\documentclass[epj]{svjour}
\usepackage{graphicx}
\usepackage{color}

\def\mb#1{\setbox0=\hbox{$#1$}\kern-.025em\copy0\kern-\wd0
\kern-0.05em\copy0\kern-\wd0\kern-.025em\raise.0233em\box0}

\begin{document}
   \title{Statistical mechanics of two-dimensional point vortices: relaxation equations and strong mixing limit}

 \author{Pierre-Henri Chavanis}

\institute{Laboratoire de Physique Th\'eorique (IRSAMC), CNRS and UPS, Universit\'e de Toulouse, F-31062 Toulouse, France\\
\email{chavanis@irsamc.ups-tlse.fr}
}

\titlerunning{Statistical mechanics of two-dimensional point vortices}

   \date{To be included later }

   \abstract{We complement the literature on the
statistical mechanics of point vortices in two-dimensional hydrodynamics. Using
a maximum entropy principle, we determine the multi-species Boltzmann-Poisson
equation and establish a form of Virial theorem. Using a maximum entropy
production principle (MEPP), we derive a set of relaxation equations towards
statistical equilibrium.  These relaxation equations can be used as a numerical
algorithm to compute the maximum entropy state. We mention the analogies with
the Fokker-Planck equations derived by Debye and H\"uckel for
electrolytes. We then consider the limit of strong mixing (or low energy). To
leading order, the relationship between the vorticity and the stream function at
equilibrium is linear and the maximization of the entropy becomes equivalent to
the minimization of the enstrophy. This expansion is similar to the
Debye-H\"uckel
approximation for electrolytes, except that the temperature is negative instead
of positive so that the effective interaction between like-sign  vortices is
attractive instead of repulsive. This leads to an organization at large scales
presenting geometry-induced phase transitions, instead of Debye shielding. We
compare the results obtained with point vortices to those obtained in the
context of the statistical mechanics of continuous vorticity fields described by
the Miller-Robert-Sommeria (MRS) theory. At linear order, we get the same
results but differences appear at the next order. In particular, the MRS theory
predicts a transition between sinh and tanh-like $\omega-\psi$ relationships
depending on the sign of ${\rm Ku}-3$ (where ${\rm Ku}$ is the Kurtosis) while there is no
such transition for point vortices which always show a sinh-like $\omega-\psi$
relationship. We derive the form of the relaxation equations in the strong
mixing limit and show that the enstrophy plays the role of a Lyapunov
functional.
\PACS{2D hydrodynamics; point vortices; kinetic theory} }

   \maketitle
%
%________________________________________________________________

\section{Introduction}

Point vortices in two-dimensional (2D) hydrodynamics provide
an interesting example of systems with long-range interactions  with application
to geophysical and astrophysical flows \cite{houches}. Furthermore, they display
many analogies (despite, of course, differences) with other systems with
long-range interactions  such as plasmas \cite{balescu}, self-gravitating
systems \cite{bt}, and the Hamiltonian mean field (HMF) model \cite{cdr}. These
systems are actively studied at present from the viewpoint
of statistical mechanics and kinetic theory.

An isolated system made of a large number $N\gg 1$ of point vortices is expected
to achieve a
statistical equilibrium state for $t\rightarrow +\infty$. As first recognized by
Onsager \cite{onsager}, this statistical equilibrium state has very peculiar
properties. Point vortices may be found at positive or negative temperatures and
behave very differently depending on the sign of the temperature.\footnote{We
emphasize that a Hamiltonian system of point vortices evolves at fixed energy,
so that the energy (not the temperature) is the relevant control parameter.
Negative temperature states correspond to high energies and positive temperature
states correspond to low energies.} At negative temperatures, point vortices of
the same sign tend to ``attract'' each other and form coherent structures
similar to the large-scale vortices observed in the atmosphere of giant planets.
 This self-organization of point vortices into coherent structures is similar to
the self-organization of stars into galaxies or globular clusters
\cite{houches}. At positive temperatures, point vortices of the same sign tend
to ``repel'' each other. In that case, the situation is similar to that of a
plasma or an electrolyte studied by Debye and H\"uckel \cite{dh1,dh2}.

The kinetic theory of point vortices is also interesting and non trivial \cite{kinonsager}. The relaxation towards the statistical equilibrium state for $t\rightarrow +\infty$ is a two-stage process.

In a first stage, the ``collisions'' (correlations) between point vortices are
negligible and the evolution of the smooth vorticity field is described by the
2D Euler equation. This equation is exact in a
proper thermodynamic limit $N\rightarrow +\infty$ in which the domain area is
fixed and the circulation of the point vortices scales as $1/N$
\cite{marchioro}. Starting from an unstable or unsteady initial
condition, the 2D Euler equation develops a complicated mixing process
and undergoes a violent relaxation towards a quasi stationary state (QSS) on the
coarse-grained scale. This
process takes place on a few dynamical times $t_D$. The QSS is a steady state of
the 2D Euler equation that depends on the initial conditions in a non-trivial
manner. Miller \cite{miller} and Robert and Sommeria \cite{rs} have proposed to
describe this QSS as a statistical equilibrium state of the 2D Euler equation.
The MRS theory is similar to the statistical theory of violent relaxation
developed by Lynden-Bell \cite{lb} in astrophysics for the Vlasov equation
describing collisionless stellar systems (see \cite{houches,csr} for a
development of the analogy between 2D vortices and stellar systems).

In a second stage, the ``collisions'' (correlations, graininess, finite $N$
effects) between the point vortices must be taken into account and the evolution
of the smooth vorticity field departs from the pure 2D Euler equation. It is
expected to be governed by a kinetic equation that relaxes for $t\rightarrow
+\infty$ towards the mean field Boltzmann distribution derived by Joyce and
Montgomery \cite{jm}. Different kinetic equations have been proposed. However,
the relaxation towards the Boltzmann distribution, and the scaling of the
relaxation time with the number $N$ of point vortices, is still a subject of
investigation (see \cite{kinonsager} and references therein). In any case, the
collisional relaxation is very slow since the relaxation time scales as $N t_D$
or may be even longer.

We note that the limits $N\rightarrow +\infty$ and $t\rightarrow +\infty$ do
not commute. If we take the $N\rightarrow +\infty$ limit before the
$t\rightarrow +\infty$ limit, the evolution of the system is described by the 2D
Euler equation that relaxes towards a non-Boltzmannian QSS described by the MRS
theory. If we take the  $t\rightarrow +\infty$ before the $N\rightarrow +\infty$
limit, the system is expected to relax towards the mean field Boltzmann
distribution described by the Joyce-Montgomery theory. For large but finite $N$,
the system quickly reaches a non-Boltzmannian QSS that is a steady state of the
2D Euler
equation, then slowly relaxes towards the Boltzmann distribution of statistical
equilibrium.

These two successive regimes of ``violent collisionless relaxation'' towards 
a non-Boltzmannian QSS and ``slow collisional relaxation'' towards the Boltzmann
distribution were predicted in \cite{pre,houches} from the kinetic theory of
point vortices.\footnote{It is sometimes argued that the MRS theory (valid for
continuous vorticity fields described by the 2D Euler equation) is a refinement
of the Joyce-Montgomery theory (valid for point vortices described by
Hamiltonian equations). This is not the case at all. In the context of point
vortices, the MRS theory and the Joyce-Montgomery theory are fundamentally
different because they apply to different regimes (collisionless relaxation vs
collisional relaxation) with very different timescales (see Appendix
\ref{sec_diff}). Therefore, they have their own domain of validity. The same
distinction applies between the Lynden-Bell distribution (describing
collisionless stellar systems) and the mean field Boltzmann distribution
(describing collisional stellar systems) in astrophysics.} They have been
confirmed and illustrated numerically in \cite{kawahara1,kawahara2}.
Self-gravitating systems and the HMF model display a similar two-stages
evolution \cite{paddy,houches,cdr}.

In this paper we consider exclusively the final statistical equilibrium state of
2D point vortices resulting from the ``collisional'' relaxation. The statistical
mechanics of 2D point vortices has been discussed in several papers, both in the
physics \cite{jm,mj,kida,pl,lp} and mathematical
\cite{caglioti1,kiessling,eyink,caglioti2,kl,sawada} literature, and a good
understanding is now achieved. In the present paper, we precise some results and
provide complements that have not been given previously. In Sec. \ref{sec_gg},
we derive the Boltzmann distribution of the multi-species point vortex gas by
using the maximum entropy principle\footnote{We generalize
previous works \cite{jm,mj,kida,pl,lp} that consider, for simplicity, only one
type of point vortices with circulation $\gamma$ or  two types of point vortices
with circulation $+\gamma$ and $-\gamma$. This generalization, although
straightforward, is important in order to compare, in Sec. \ref{sec_sm},  the
limit of strong mixing for point vortices and continuous vorticity fields.}
(this variational problem is justified
from the theory of large deviations in Appendix
\ref{sec_heur}). We discuss the proper thermodynamic limit of the point vortex
gas, the notion of ensemble inequivalence, and the Virial theorem. We also
stress some limitations of the celebrated sinh-Poisson equation. In Sec.
\ref{sec_mepp}, we derive a set of relaxation equations towards the statistical
equilibrium state by using a maximum entropy production principle
(MEPP).\footnote{The MEPP was introduced in the context of the
MRS theory \cite{rsmepp} for continuous vorticity fields. We apply it here to
the point vortex gas in order to see the similarities and the differences.} We
consider the multi-species case and take all the constraints into
account (circulation of each species, energy, angular momentum, and linear
impulse). We simplify these equations for the globally neutral two-species
system and for the single species system. We mention the connections (as well as
the differences) between these relaxation equations and the Fokker-Planck
equations derived by Debye and H\"uckel \cite{dh2} in their theory of
electrolytes. We also discuss the analogy with the Smoluchowski-Poisson system
describing self-gravitating Brownian particles \cite{sopik} and with the
Keller-Segel model describing the chemotaxis of bacterial populations \cite{ks}.
In Sec. \ref{sec_sm} we consider a limit of strong mixing (or low energy) and
simplify the equations of the statistical theory. To leading order, the
$\omega-\psi$ relationship between the smooth vorticity and the stream function
becomes linear and the maximization of the entropy becomes equivalent to the
minimization of the enstrophy. This expansion is similar to the Debye-H\"uckel
approximation for electrolytes, except that the temperature is negative instead
of positive so that the effective interaction between like-sign  vortices is
attractive instead of repulsive. This leads to an organization at large scales
presenting geometry-induced phase transitions, instead of Debye shielding. We
mention the connection with the phenomenological minimum enstrophy principle
\cite{leith} and with the phenomenon of ``condensation'' put forward by
Kraichnan \cite{kraichnan} in his statistical theory of 2D turbulence in
spectral space. We also compare the results obtained for point vortices with
those obtained by Chavanis and Sommeria \cite{jfm1} in the context of the
statistical mechanics of continuous vorticity fields described by the
Miller-Robert-Sommeria (MRS) theory. At linear order, we get the same results
but differences appear at the next order. In particular, the MRS theory predicts
a transition between sinh and tanh-like $\omega-\psi$ relationships depending on
the sign of ${\rm Ku}-3$ (where ${\rm Ku}$ is the Kurtosis) while there is no such
transition for point vortices which always show a sinh-like $\omega-\psi$
relationship (this important difference is detailed in 
Appendices \ref{sec_smpv} and \ref{sec_smcv}). Finally, we derive 
the form of the relaxation equations in the
strong mixing limit and show that the enstrophy plays the role of a Lyapunov
functional.

\section{Statistical equilibrium state of a multi-species point vortex gas}
\label{sec_gg}

\subsection{The Hamiltonian equations}
\label{sec_s}

The dynamical evolution of a multi-species system of point vortices in
two dimensions is described by the Hamiltonian equations
\cite{newton}:
\begin{eqnarray}
\gamma_{i}{dx_{i}\over dt}={\partial H\over\partial y_{i}}, \qquad \gamma_{i}{dy_{i}\over dt}=-{\partial H\over\partial x_{i}},
\label{s1}
\end{eqnarray}
\begin{eqnarray}
H=-{1\over 2\pi}\sum_{i<j}\gamma_{i}\gamma_{j}\ln |{\bf r}_{i}-{\bf r}_{j}|,
\label{s2}
\end{eqnarray}
where $\gamma_{i}$ is the circulation of point vortex $i$. For simplicity, we have written the Hamiltonian
in an infinite domain. In a bounded domain, it has to
be modified so as to take into account the contribution of vortex
images.

A particularity of
this Hamiltonian system, first noticed by Kirchhoff
\cite{kirchhoff}, is that the coordinates $(x,y)$ of the point
vortices are canonically conjugate. Another particularity of the point vortex
model is that the Hamiltonian (\ref{s2}) does not possess the usual
kinetic term $\sum_{i}m_{i}v_{i}^{2}/2$ present for
material particles.  This is because point vortices have no
inertia. Therefore, a point vortex produces a velocity, not an
acceleration (or a force), contrary to other systems of particles in
interaction like electric charges in a plasma \cite{balescu} or stars
in a galaxy \cite{bt}. In a sense, this corresponds to the conception of motion according to Descartes in contrast to Newton.

The point vortex system conserves the energy $E=H$ and the number $N_{a}$
of vortices of each species $a$ (this is equivalent to the total circulation of each species
$\Gamma_{a}=N_{a}\gamma_{a}$). There are additional conserved quantities depending
on the geometry of the domain. The angular momentum
$L=\sum_{i}\gamma_{i} r_{i}^{2}$ is conserved in an infinite domain and
in a disk due to the invariance by rotation. The linear impulse ${\bf P}=\sum_{i}\gamma_{i}{\bf
r}_{i}$ is conserved in an infinite domain and in a channel due to the invariance
by translation.

\subsection{The Onsager theory  and the different approach of Ruelle}
\label{sec_or}

Assuming ergodicity, the point vortex gas is expected to achieve a
statistical equilibrium state for $t\rightarrow +\infty$. The
statistical mechanics of point vortices is very peculiar and was first
discussed by Onsager \cite{onsager} in a bounded domain.

At sufficiently large
energies, point vortices of the same sign tend to
``attract'' each other and group themselves in ``clusters'' or
``supervortices'' similar to the large-scale vortices observed in the
atmosphere of giant planets (e.g. Jupiter's great red spot). When all the vortices
have the same sign, this leads to monopoles (cyclones or anticyclones)
and when vortices have positive and negative signs, this leads
to dipoles (a pair of cyclone/anticyclone) or even tripoles.  These
organized states are characterized by {\it negative}
temperatures.\footnote{Fundamentally, the
microcanonical temperature $T(E)$ is defined by $\beta(E)=1/T(E)=dS/dE$ where
$S(E)=\ln g(E)$ is the entropy and $g(E)=\int \delta(H-E)\,
dx_{1}dy_{1}...dx_{N}dy_{N}$ is the density of states with energy $E$. The
density of states is related to the structure function $\Phi(E)=\int_{H\le
E}dx_{1}dy_{1}...dx_{N}dy_{N}$ by $g(E)=d\Phi/dE$. For point vortices, the
structure function is finite for $E\rightarrow +\infty$ because the phase space
coincides with the
configuration space ($\Phi(+\infty)=\int
dx_{1}dy_{1}...dx_{N}dy_{N}=V^{N}$ where $V$ is the area of the
system). Therefore, $g(E)\rightarrow 0$ for $E\rightarrow +\infty$ so there must
exist an energy $E_c$ above which the entropy decreases, implying negative
temperatures \cite{onsager}. These negative temperature states have high
energies so they correspond to large-scale structures where like-sign point 
vortices group themselves in clusters. On the other hand, for $E<E_c$, the
entropy increases with the energy, implying positive temperatures. These
positive temperature states have low
energies so they correspond to disorganized states where like-sign point
vortices repell each other.} The existence of negative temperatures for point
vortices
should not cause surprise. For material particles, the temperature is
a measure of the velocity dispersion and it must be positive. Indeed,
it appears in the Boltzmann factor $e^{-\beta v^2/2}$ which can be normalized only for
$\beta>0$. However, since point vortices have no inertia, there is no
such term in the equilibrium distribution of point vortices (see, e.g., Eq. (\ref{m3}) below)
and the temperature can be negative.

At sufficiently large negative energies,  like-sign vortices tend to
``repel'' each other. This corresponds to positive
temperature states. When all the vortices of the system have the
same sign, they accumulate on the boundary of the domain. Alternatively, when
the system is neutral, opposite-sign vortices tend to
``attract'' each other resulting in a spatially homogeneous
distribution with strong correlations between the vortices. In that case,
the point vortex gas is similar to a Coulombian plasma. This is the
situation considered by Fr\"ohlich and Ruelle \cite{fr}. As discussed by
Ruelle \cite{ruelle}, we may expect a phase transition (related to the
Kosterlitz-Thouless transition) as a function of the temperature.  At
large positive temperatures, the system is in a ``conducting phase''
(in the plasma analogy) with free vortices that can screen other vortices.
In that case, there is Debye shielding like for a
Coulombian plasma. At low positive temperatures, opposite-sign
vortices tend to form pairs $(+,-)$ and the gas is in a ``dielectric
phase'' where all the charges are bound, forming dipolar pairs. These
``dipoles'' are similar to ``atoms'' $(+e,-e)$ in plasma physics. In
that case, there is no screening.

In summary, at negative temperatures, point vortices
of the same sign tend to ``attract'' each other and form large-scale structures.
This is similar to the case of self-gravitating
systems, interacting by Newtonian gravity, in astrophysics. By contrast,  at
positive temperatures,  point vortices of the same sign tend to ``repel'' each
other, and opposite-sign vortices tend to ``attract'' each others. This is
similar to the case of electric charges, interacting by the Coulombian
force, in  plasma physics and in electrolytes.

\subsection{The thermodynamic limit and the mean field approximation}
\label{sec_mfap}

To obtain more quantitative results, Joyce and Montgomery \cite{jm} and
Lundgren and Pointin \cite{lp} have considered a mean field
approximation. For a single
species system of point vortices, this is valid in a
proper thermodynamic limit
$N\rightarrow +\infty$ in such a way that the normalized energy
$\epsilon=E/N^{2}\gamma^{2}$, the normalized entropy $s=S/N$,  and the
normalized temperature
$\eta=\beta N\gamma^{2}$ are of order unity. On physical grounds, it is reasonable to
consider that the area $V$ of the domain and the total circulation $\Gamma=N\gamma$ of the
vortices are of order unity. Then, by rescaling the parameters appropriately,
the  thermodynamic limit of the spatially inhomogeneous point vortex gas
corresponds to $N\rightarrow +\infty$
with $\gamma\sim 1/N$ and $V\sim 1$. In that case,  $E\sim 1$, $S\sim N$, and
$\beta\sim N$. Since the coupling constant $\beta\gamma^{2}=\eta/N\sim 1/N$ 
goes to zero for $N\rightarrow +\infty$, we are in a {\it weak coupling} limit.
For $N\rightarrow +\infty$, the correlations between point vortices can be
neglected and the mean field approximation becomes exact
\cite{caglioti1,kiessling,eyink,caglioti2,kl,sawada}. The statistical
equilibrium state is spatially inhomogeneous and is described by the one-body
distribution $P_1({\bf r})$ or by the smooth vorticity field $\omega({\bf
r})=N\gamma P_1({\bf r})$ determined by the Boltzmann-Poisson equation. These
results can be generalized to a spatially inhomogeneous multi-species system of
point vortices. In that case, the system is dominated by the mean flow.

However, for a spatially homogeneous neutral system of positive and
negative point vortices with circulation $\pm\gamma$, the situation is
different. In that case, the one-body distribution  $P_1({\bf r})=1$ is trivial
(no mean flow) and the physics of the problem is controlled by the two-body
correlation
function $P'_2({\bf r},{\bf r}')$. The relevant thermodynamic limit is the usual
thermodynamic limit  $N,V\rightarrow +\infty$ with $N/V$ fixed (and
$\gamma\sim 1$). At high positive
temperatures, there is Debye shielding like in plasma physics and the
correlation function tends to zero exponentially rapidly. This is the situation
considered by Fr\"ohlich and Ruelle \cite{fr}, and Ruelle \cite{ruelle}. At
negative temperatures, there is ``anti-shielding'' and the vortices of the same
sign tend to form clusters as investigated by Edwards and Taylor \cite{et} (see
also \cite{pvbrownian,virialvortex}). The two-body correlation
function of a spatially homogeneous neutral system of point vortices has
been
determined by Pointin and Lundgren \cite{pl}   in a
periodic domain and their approach has been recently generalized  by Esler {\it
et al.} \cite{esler} in arbitrary domains. Their results are
valid in a proper thermodynamic limit $N\rightarrow +\infty$ in such a way that
the normalized energy $\epsilon'=E/N\gamma^{2}$, the entropy $S$, and the
normalized temperature $\eta=\beta N\gamma^{2}$ are of order unity. By
rescaling the parameters appropriately, the thermodynamic limit of the
spatially homogeneous neutral point vortex gas corresponds to $N\rightarrow
+\infty$ with $\gamma\sim \pm 1/N$, $V\sim 1$,  $E\sim 1/N$, $S \sim 1$, and
$\beta\sim N$.

In this paper, we exclusively consider the case of spatially inhomogeneous systems described by the Boltzmann-Poisson equation \cite{jm,lp}. We call $\omega_a({\bf r},t)=\gamma_a n_a({\bf r},t)$ the vorticity of species $a$ and $\omega({\bf r},t)=\sum_a \omega_a({\bf r},t)$ the total vorticity ($n_a$ is the density of point vortices of species $a$ and $n=\sum_a n_a$ is the total density). The vorticity is related to the stream function by the Poisson equation
\begin{eqnarray}
-\Delta\psi=\omega=\sum_{a}\omega_{a}.
\label{mf1}
\end{eqnarray}
The velocity field is given by ${\bf u}=-{\bf z}\times \nabla\psi$. In the mean field approximation, the conservation laws can be written as
\begin{equation}
\label{mf2}
E={1\over 2}\int {\omega}\psi \, d{\bf r},\quad \Gamma_{a}=\int \omega_a\, d{\bf r},
\end{equation}
\begin{equation}
\label{mf4}
L=\int {\omega} r^2\, d{\bf r},\quad {\bf P}=-{\bf z}\times \int {\omega} {\bf r} \, d{\bf r}.
\end{equation}

\subsection{The Boltzmann entropy}

In order to determine the statistical
equilibrium state of a system of point vortices with different
circulations, we generalize the maximum entropy approach of Joyce and Montgomery
\cite{jm}. Following the Boltzmann procedure, we divide the
domain into a very large number of microcells with size $h$
(ultimately $h\rightarrow 0$).  A microcell can be occupied by an
arbitrary number of point vortices.  We now group these
microcells into macrocells each of which contains many microcells but
remains nevertheless small compared to the extension of the whole
system. We call $\nu$ the number of microcells in a macrocell. A
microstate is specified by  precise position $\lbrace {\bf r}_1,...,{\bf r}_N\rbrace$
of the point vortices while a macrostate is specified by the number $\lbrace n_{ia}
\rbrace$ of point vortices of each species in each macrocell
(irrespective of their position in the cell). Using a
combinatorial analysis, the number of microstates
corresponding to the macrostate
$\lbrace n_{ia} \rbrace$ is
\begin{eqnarray}
W(\lbrace n_{ia} \rbrace )=\prod_{a}N_{a}!\prod_{i}\frac{\nu^{n_{ia}}}{n_{ia}!}.
\label{b1}
\end{eqnarray}
We introduce $n_{a}({\bf r})$ the smooth density of point vortices of
species $a$ at ${\bf r}$. The smooth vorticity of species $a$ is
then $\omega_{a}({\bf r})=\gamma_{a} n_{a}({\bf r})$. If we define the
Boltzmann entropy by $S=\ln W$, use the Stirling formula for
$n_{ia}\gg 1$, and take the continuum limit, we get
\begin{eqnarray}
S[\omega_a]=-\sum_{a}\int {\omega_{a}\over \gamma_{a}}\ln {\omega_{a}\over N_a\gamma_{a}} \ d{\bf r}.
\label{b2}
\end{eqnarray}
The Boltzmann entropy (\ref{b2}) measures the number of microstates $W[\omega_a]$
corresponding to the macrostate  $\lbrace \omega_{a}({\bf
r})\rbrace$. At statistical equilibrium, the system is expected
to be in the {\it most probable} macrostate, i.e. the one that is the most
represented at the microscopic level.

\subsection{The microcanonical ensemble}

Since the energy is conserved, the proper statistical ensemble is the
microcanonical ensemble. In the microcanonical ensemble, all the accessible
microstates (those satisfying all the constraints) are {\it equiprobable} at
statistical equilibrium. This is the fundamental postulate of
statistical mechanics, assuming ergodicity. Therefore, the probability of an
accessible macrostate is proportional to $W[\omega_a]=e^{S[\omega_a]}$. As a
result, the most probable macrostate is obtained by maximizing the Boltzmann
entropy (\ref{b2}) while conserving the mean field energy (\ref{mf2}-a), the
circulation of each species (\ref{mf2}-b), the angular momentum (\ref{mf4}-a),
and the impulse (\ref{mf4}-b).

We thus have to solve the maximization problem (see Appendix \ref{sec_heur}):
\begin{eqnarray}
\label{m1}
S(E,\Gamma_a,L,{\bf P})=\max_{\omega_a}\quad \lbrace S[\omega_a]\quad | \quad E, \, \Gamma_a, \, L, \, {\bf P} \quad {\rm fixed} \rbrace.\nonumber\\
\end{eqnarray}
The critical points of the Boltzmann entropy at fixed $E$, $\Gamma_a$, $L$ and ${\bf P}$ are determined by the condition
\begin{equation}
\label{m2}
\delta
S-\beta\delta E-\sum_a \alpha_a \delta\Gamma_a-\beta\frac{\Omega}{2}\delta L+\beta {\bf U}\cdot \delta {\bf P}=0,
\end{equation}
where $\beta$ (inverse temperature), $\alpha_a$ (chemical potentials), $\Omega$ (angular velocity), and ${\bf U}$ (linear velocity) are appropriate Lagrange
multipliers. Using
\begin{equation}
\delta S=-\sum_a\int\frac{\delta\omega_a}{\gamma_a}\left (\ln \frac{\omega_a}{N_a\gamma_a}+1\right )\, d{\bf r},
\end{equation}
\begin{equation}
\delta E=\int\psi\delta\omega\, d{\bf r},\qquad \delta\Gamma_a=\int \delta\omega_a\, d{\bf r},
\end{equation}
\begin{equation}
\delta L=\int\delta\omega r^2\, d{\bf r},\qquad \delta {\bf P}=-{\bf z}\times \int \delta\omega {\bf r}\, d{\bf r},
\end{equation}
we obtain the mean field Boltzmann distribution for each species\footnote{In an
unbounded domain, we must consider vortices of the same circulation otherwise
they would form pairs (dipoles) and ballistically escape to infinity, implying
an absence of statistical equilibrium. Furthermore, due to the
$e^{-\beta\gamma_a\Omega r^2/2}$ factor, the distribution (\ref{m3}) can
possibly be normalizable only if ${\rm sgn}(\gamma)\beta\Omega\ge 0$.}
\begin{equation}
\label{m3}
\omega_a=\gamma_{a}n_{a}=A_a \gamma_a e^{-\beta\gamma_a\psi_{eff}},
\end{equation}
where $A_{a}=N_a e^{-1-\alpha_{a}\gamma_{a}}>0$ is determined by the
normalization condition (conservation of the number of point vortices of each
species) yielding
\begin{equation}
\label{m3b}
A_a=\frac{N_{a}}{\int e^{-\beta\gamma_a\psi_{eff}}\, d{\bf r}}.
\end{equation}
The total vorticity is then given by
\begin{eqnarray}
\label{m4}
\omega=\sum_a A_a \gamma_a e^{-\beta\gamma_a\psi_{eff}}=-\frac{1}{\beta}\frac{dn}{d\psi_{eff}}=f_{\alpha,\beta}(\psi_{eff}).
\end{eqnarray}
In these expressions $\psi_{eff}$ is the relative stream function
\begin{eqnarray}
\psi_{eff}=\psi+\frac{\Omega}{2}r^2-{\bf U}_{\perp}\cdot {\bf r},
\label{m5}
\end{eqnarray}
where ${\bf U}_{\perp}={\bf z}\times {\bf U}$. This corresponds to a relative
velocity field ${\bf u}_{eff}={\bf u}-{\bf \Omega}\times {\bf r}-{\bf U}$.

Differentiating Eq. (\ref{m4}) with respect to $\psi_{eff}$, we obtain the identity
\begin{equation}
\label{m6}
\omega'(\psi_{eff})=-\beta\omega_{2}(\psi_{eff}),
\end{equation}
where
\begin{equation}
\label{m7}
\omega_2\equiv \sum_a\gamma_a\omega_a=\sum_a\gamma_a^2 n_a\ge 0
\end{equation}
is the local microscopic enstrophy. Since ${\omega}={\omega}(\psi_{eff})$, the statistical
theory predicts that the vorticity ${\omega}({\bf r})$ is a {stationary solution} of the 2D
Euler equation\footnote{The 2D Euler equation describes the dynamical evolution
of the smooth vorticity field $\omega({\bf r},t)$ of the point vortex gas in the
limit $N\rightarrow +\infty$ \cite{kinonsager,marchioro}. It is analogous to the
Vlasov equation in plasma physics \cite{balescu} and stellar dynamics
\cite{bt}.} in a frame rotating with angular velocity $\Omega$ and/or
translating with a constant velocity ${\bf U}$ (see Appendix \ref{sec_euler}).
Explicit examples of such structures are given in \cite{jfm2}. On the other
hand, according to Eqs. (\ref{m6}) and (\ref{m7}), the ${\omega}-\psi_{eff}$
relationship is a {monotonic}
function that is increasing at {negative temperatures} $\beta<0$ and
decreasing at positive temperatures $\beta>0$.

Substituting Eq. (\ref{m4}) in the Poisson equation
(\ref{mf1}), the equilibrium stream function is obtained by solving
the multi-species Boltzmann-Poisson equation
\begin{eqnarray}
-\Delta\psi=\sum_{a}A_{a}\gamma_a e^{-\beta\gamma_{a}\psi_{eff}}=f_{\alpha,\beta}(\psi_{eff})
\label{m8}
\end{eqnarray}
with adequate boundary conditions, and relating the Lagrange
multipliers $\beta$, $\alpha_a$, $\Omega$, and ${\bf U}$ to the constraints $E$,
$\Gamma_a$, $L$, and ${\bf P}$. We also have to make sure that the distribution
(\ref{m3}) is a (local) maximum of entropy, not a minimum or a
saddle point. One can show \cite{stabvortex} that
the Boltzmann distribution (\ref{m3}) is a (local) maximum
of constrained entropy if, and only if,
\begin{eqnarray}
\delta^{2}J\equiv -\frac{1}{2}\sum_{\alpha}\int \frac{(\delta \omega_{\alpha})^{2}}{\gamma_{\alpha}\omega_{\alpha}} \, d{\bf r}-\frac{\beta}{2}\int \delta {\omega}\delta\psi  \, d{\bf r}<0\nonumber\\
\forall \delta\omega_a \quad|\quad \delta\Gamma_a=\delta L=\delta {\bf P}=\delta E=0,\qquad\qquad
\label{pvm7}
\end{eqnarray}
i.e. for all perturbations that conserve the circulations of each
species, the angular momentum, the linear impulse, and the energy at first order. This is the criterion of thermodynamical stability
in the microcanonical ensemble.

We note that the vorticity profiles of
different species of point vortices are related to each other by
\begin{eqnarray}
\left ({\omega_{a}\over A_{a}\gamma_a}\right )^{1/\gamma_{a}}=\left
({\omega_{b}\over A_{b}\gamma_b}\right )^{1/\gamma_{b}}
\label{gg11}
\end{eqnarray}
hence
\begin{eqnarray}
\omega_{a}({\bf r})=C_{ab}|\omega_{b}({\bf r})|^{\gamma_{a}/\gamma_{b}},
\label{gg12}
\end{eqnarray}
where $C_{ab}$ is independent on the position. Assuming that
$\gamma_{a}>0$, $\gamma_{b}>0$ and that $\omega_{b}({\bf r})$
decreases with the distance (which corresponds to equilibrium states
with $\beta<0$), this relation indicates that intense vortices
($\gamma_{a}>\gamma_{b}$) are more concentrated at the center, on
average, than weaker vortices. On the other hand, Eq. (\ref{gg12})
shows that opposite-sign vortices tend to separate (at negative
temperatures). More generally,
Eq. (\ref{gg12}) characterizes the segregation between vortices with
different circulations.

{\it Remark:} The mean field Boltzmann distribution (\ref{m3}) was first derived by Joyce and Montgomery \cite{jm} from a maximum entropy principle and by Lundgren and Pointin \cite{lp}
from the Yvon-Born-Green (YBG) hierarchy (see also the discussion in \cite{virialvortex}). A rigorous derivation of the mean field
equations is given in \cite{caglioti1,kiessling,eyink,caglioti2,kl,sawada}. We
stress that
the statistical equilibrium state relies on an hypothesis of ergodicity (or at
least efficient mixing) assuming that all the accessible microstates are
equiprobable. There is no guarantee that the evolution of the point
vortex gas  is ergodic in all cases (on the contrary, there are strong arguments
that it is not ergodic) and, therefore, the relaxation towards the Boltzmann
distribution remains an open problem. To determine whether the point vortex gas
truly relaxes towards the Boltzmann distribution, and to establish the scaling
of the relaxation time with $N$,  we must develop a kinetic theory of point
vortices \cite{kinonsager}.

\subsection{The canonical ensemble}
\label{sec_cpv}

In the canonical ensemble, where $\beta$, $\Omega$ and ${\bf U}$ are assumed to be fixed, the probability of a microstate with energy $H$ is proportional to the Boltzmann factor $e^{-\beta H_{eff}}$ where $H_{eff}=H+\frac{\Omega}{2} L- {\bf U}\cdot {\bf P}$ is the effective Hamiltonian in the moving frame where the flow is steady. Therefore, the probability of an accessible  macrostate is proportional to $W[\omega_a] e^{-\beta E_{eff}[\omega]}=e^{S[\omega_a]-\beta E_{eff}[\omega]}=e^{J[\omega_a]}$. As a result, the most probable macrostate is obtained by maximizing the Boltzmann free energy (more properly the Massieu function):
\begin{eqnarray}
\label{c2}
J[\omega_{\alpha}]=S[\omega_{\alpha}] -\beta
E[\omega]-\beta\frac{\Omega}{2} L[\omega]+\beta {\bf U}\cdot {\bf P}[\omega]
\end{eqnarray}
while conserving the circulation of each species (\ref{mf2}-b).

We thus have to solve the
maximization problem (see Appendix \ref{sec_heur}):
\begin{eqnarray}
\label{c1}
J(\beta,\Gamma_a,\Omega,{\bf U})=\max_{\omega_a}\quad \lbrace J[\omega_{\alpha}]\quad | \quad \Gamma_a \quad {\rm fixed} \rbrace.
\end{eqnarray}
The critical points of the Boltzmann free energy at fixed $\Gamma_a$ are determined by the condition
\begin{equation}
\label{c3}
\delta
J-\sum_a \alpha_a \delta\Gamma_a=0,
\end{equation}
where  $\alpha_a$ (chemical potentials) are appropriate Lagrange
multipliers. This leads to the mean field Boltzmann distribution (\ref{m3}) as in the microcanonical ensemble. The Boltzmann distribution (\ref{m3}) is a (local) maximum
of constrained free energy $J$ if, and only if,
\begin{eqnarray}
\delta^{2}J\equiv -\frac{1}{2}\sum_{\alpha}\int \frac{(\delta \omega_{\alpha})^{2}}{\gamma_{\alpha}\omega_{\alpha}} \, d{\bf r}-\frac{\beta}{2}\int \delta {\omega}\delta\psi  \, d{\bf r}<0\nonumber\\
\forall \delta\omega_a \quad|\quad \delta\Gamma_a=0,\qquad\qquad\qquad
\label{c4}
\end{eqnarray}
i.e. for all perturbations that conserve the circulations of each
species: $\delta\Gamma_a=0$. This is the criterion of thermodynamical stability
in the canonical ensemble.

\subsection{Generic ensemble inequivalence for systems with long-range interactions}
\label{sec_mepeq}

The variational problems (\ref{m1}) and (\ref{c1})  have the 
same critical points. However, these variational problems may not be equivalent.
The set of solutions of (\ref{m1}) may not coincide with the set of solutions of
(\ref{c1}). It can be shown that a solution of a variational problem is always
the solution of a more constrained dual variational problem \cite{ellis}.
Therefore, a solution of (\ref{c1}) with given $\beta$, $\Omega$, and ${\bf U}$
is always a solution of (\ref{m1}) with the corresponding $E$, $L$, and ${\bf
P}$. Thus, canonical stability implies microcanonical stability\footnote{This
can be checked at the level of the second order variations. Indeed, if
inequality (\ref{c4}) is satisfied for all perturbations that conserve the
circulations of each species (canonical stability criterion), it is {\it a
fortiori} satisfied for all perturbations that conserve in addition the angular
momentum, the linear impulse, and the energy at first order (microcanonical
stability criterion).}: $(\ref{c1}) \Rightarrow (\ref{m1})$. However, the
converse is wrong: a solution of (\ref{m1}) is not necessarily a solution of
(\ref{c1}). When this happens, we speak of ensemble inequivalence. Ensemble
inequivalence is generic for systems with long-range interactions but it is not
compulsory.\footnote{For self-gravitating systems, there exist a
region of ensemble inequivalence in a range of energies in which the
configurations have negative specific heats (these states are accessible in the
microcanonical ensemble but not in the canonical ensemble) \cite{ijmpb}. For the
Hamiltonian Mean Field (HMF) model, the ensembles are equivalent for all
energies even though the particles have long-range interactions \cite{cdr}. For
2D vortices, the equivalence or inequivalence of the statistical ensembles 
depends on the shape of the domain and on the value of the control parameters
$\Gamma_a$, $E$, $L$ and ${\bf P}={\bf 0}$ in a non-trivial manner (one has to
solve the equations of the statistical theory and study the variational problems
(\ref{m1}) and (\ref{c1}) specifically). Note that the stability of the
solutions in each ensemble (and consequently the possible occurrence of ensemble
inequivalence) may be decided by plotting the series of equilibria and using the
Poincar\'e theorem (see, e.g., \cite{stabvortex,ijmpb} for more details).}

For isolated systems that evolve at fixed energy, like the point vortex model, the proper statistical ensemble is the microcanonical ensemble. Since the energy is non-additive, the canonical ensemble is not physically justified to describe a subpart of the system (contrary to systems with short-range interactions) \cite{cdr}. However, since the canonical variational problem (\ref{c1}) provides a {\it sufficient} condition of microcanonical stability it can be useful in that respect. Indeed, if we can show that the system is canonically stable, then it is granted to be microcanonically stable. Therefore, we can start by studying the canonical stability problem, which is simpler, and consider the microcanonical stability problem only if the canonical ensemble does not cover the whole range of parameters ($E$, $\Gamma_a$, $L$, ${\bf P}$).

{\it Remark:} the canonical ensemble is mathematically justified for a system of ``Brownian point vortices'' \cite{pvbrownian} described by $N$ coupled stochastic Langevin equations (instead of Hamiltonian equations). However, this system is essentially academic and it is not clear whether it can have physical applications.

\subsection{The Virial theorem}
\label{sec_virial}

We can derive a form of Virial theorem for point vortices at statistical equilibrium \cite{virialvortex}. Taking the logarithmic derivative of Eq. (\ref{m3}), we obtain
\begin{eqnarray}
\label{vv1}
\nabla\omega_a=-\beta \gamma_a\omega_a\nabla\psi_{eff}
\end{eqnarray}
with
\begin{equation}
\label{vmf1b}
\nabla\psi_{eff}=\nabla\psi+\Omega {\bf r}-{\bf U}_{\perp}.
\end{equation}
Dividing Eq. (\ref{vv1}) by $\gamma_a$, summing over the species, and introducing the local ``pressure'' $p({\bf r})=n({\bf r})T=\sum_a (\omega_a({\bf r})/\gamma_a) T$,  we get
\begin{equation}
\label{vmf1}
\nabla p+\omega\nabla\psi_{eff}={\bf 0}.
\end{equation}
This equation is
similar to the condition of hydrostatic equilibrium for self-gravitating systems
in a rotating frame \cite{bt}. The pressure is positive at positive temperatures and
negative at negative temperatures. Taking the scalar product of Eq. (\ref{vmf1})
with ${\bf r}$, integrating over the entire domain, and integrating by parts the
pressure term, we obtain the Virial theorem
\begin{eqnarray}
\label{vmf2}
2\int p\, d{\bf r}-{\cal V}-\Omega L+{\bf U}\cdot {\bf P}=2PV,
\end{eqnarray}
where
\begin{eqnarray}
\label{mfiv1}
{\cal V}=\int\omega {\bf r}\cdot \nabla\psi\, d{\bf r}
\end{eqnarray}
is the Virial of the point vortex gas, $V$ is the area of the domain, and  
$P=\frac{1}{2V}\oint p {\bf r}\cdot d{\bf S}$ is the average pressure on the
boundary of the domain (if the pressure is uniform on the boundary, with value
$p_b$, then $P=p_b$). We note that the angular momentum $L=\int\omega r^2\,
d{\bf r}$ of point vortices is similar to the moment of inertia of material
particles. Using the isothermal equation of state $p({\bf r})=n({\bf r})T$, we
can rewrite the Virial theorem as
\begin{eqnarray}
\label{vmf3}
2N T-{\cal V}-\Omega L+{\bf U}\cdot {\bf P}=2PV.
\end{eqnarray}

In a domain without special symmetry, $L$ and ${\bf P}$ are not conserved so that $\Omega=0$ and ${\bf U}={\bf 0}$. In that case, the foregoing expression reduces to
\begin{eqnarray}
\label{vmf3b}
2N T-{\cal V}=2PV.
\end{eqnarray}

In an unbounded domain, and for axisymmetric flows in a disk of radius $R$, the Virial ${\cal V}$ is given by \cite{virialvortex}:
\begin{eqnarray}
\label{mfiv3}
{\cal V}=-\frac{\Gamma^{2}}{4\pi}.
\end{eqnarray}
In that case, the Virial theorem can be rewritten as\footnote{In an unbounded domain, we can impose ${\bf P}={\bf 0}$ by taking the origin at the center of vorticity.  In a disk the linear impulse is not conserved so that ${\bf U}={\bf 0}$.}
\begin{eqnarray}
\label{vmf4}
PV=N(T-T_{c})-\frac{1}{2}\Omega L
\end{eqnarray}
with $P=p(R)=n(R)T$ and $V=\pi R^2$. We have introduced the critical temperature
\begin{equation}
\label{vmf5}
T_{c}=-\frac{\Gamma^{2}}{8\pi N}.
\end{equation}
For a single species system of point vortices, we have
\begin{equation}
\label{vmf6}
T_{c}=-\frac{N\gamma^{2}}{8\pi}.
\end{equation}
The Virial theorem (\ref{vmf4})
relates the angular velocity $\Omega$ to the angular momentum $L$, the
temperature $T$, and the pressure $P$ on the boundary. When $\Omega=0$, it reduces to
\begin{eqnarray}
\label{vmf4bis}
PV=N(T-T_{c}).
\end{eqnarray}
On the other hand, in an unbounded domain, using $PV\rightarrow 0$ at large distances, we obtain the identity
\begin{eqnarray}
\label{vmf4b}
\frac{1}{2}\Omega L=N(T-T_{c}).
\end{eqnarray}
Additional results and discussion are given in \cite{virialvortex}.

\subsection{The sinh-Poisson equation}
\label{sec_sinh}

If we consider a globally neutral system of point vortices with $N/2$ point vortices of circulation $\gamma$ and  $N/2$ point vortices of circulation $-\gamma$, we find that the Boltzmann-Poisson equation (\ref{m8}) takes the form
\begin{eqnarray}
\label{sinh1}
\omega=-\Delta\psi=\gamma\left (A_+
e^{-\beta\gamma\psi_{eff}}-A_{-}e^{\beta\gamma\psi_{eff}}\right )
\end{eqnarray}
with
\begin{eqnarray}
\label{sinh2}
A_{\pm}=\frac{N}{2\langle e^{\mp\beta\gamma\psi_{eff}}\rangle},
\end{eqnarray}
where the brackets denote a domain average. In a bounded domain, this equation
has a non-trivial solution $\psi\neq 0$ only if $\beta<\beta_*<0$
\cite{kiessling95}. If we assume that $A_+=A_{-}=A$ (which corresponds to the
equality of the chemical potential of the two species), we obtain the celebrated
sinh-Poisson equation \cite{jm}:
\begin{eqnarray}
\label{sinh3}
\omega=-\Delta\psi=2\gamma A \sinh (-\beta\gamma\psi_{eff}).
\end{eqnarray}
However, there is no fundamental reason why $A_+=A_{-}$ so this assumption is
restrictive. For example, Eq. (\ref{sinh3}) 
holds if $\langle \psi_{eff}^{2n+1}\rangle=0$ for all $n\in \mathbf{N}$ but this
condition is not always satisfied. In particular, some important solutions
may be forgotten by using the sinh-Poisson equation (\ref{sinh3}) instead of the
Boltzmann-Poisson equation (\ref{sinh1}), as discussed in Sec. \ref{sec_sm}.

\section{Relaxation equations based on a Maximum Entropy Production Principle}
\label{sec_mepp}

In this section, using a MEPP, we derive a set of equations for a multi-species point vortex gas that relax towards statistical equilibrium.

\subsection{Relaxation equations for a multi-species point vortex gas}
\label{sec_multimepp}

The conservation of the number of point vortices of each species can be expressed in the local form by an equation of the form
\begin{eqnarray}
\frac{\partial\omega_{a}}{\partial t}+{\bf u}\cdot \nabla\omega_{a}=-\nabla\cdot {\bf J}_{a},
\label{mepp1}
\end{eqnarray}
where ${\bf J}_{a}({\bf r},t)$ is the current of species $a$ to be determined. The evolution of the total vorticity $\omega=\sum_{a}\omega_{a}$ is given by
\begin{eqnarray}
\frac{\partial\omega}{\partial t}+{\bf u}\cdot \nabla\omega=-\nabla\cdot {\bf J},
\label{mepp2}
\end{eqnarray}
where ${\bf J}=\sum_{a}{\bf J}_{a}$ is the total current of vorticity.

We can express the time variation of energy, angular momentum, and linear impulse in terms of ${\bf J}$ using Eqs. (\ref{mf2}-a), (\ref{mf4}), and (\ref{mepp2}). This leads to the constraints
\begin{eqnarray}
\dot E=\int {\bf J}\cdot\nabla\psi\, d{\bf r}=0,
\label{mepp3}
\end{eqnarray}
\begin{eqnarray}
\dot L=2\int {\bf J}\cdot {\bf r}\, d{\bf r}=0,
\label{mepp4}
\end{eqnarray}
\begin{eqnarray}
\dot {\bf P}=-{\bf z}\times \int {\bf J}\, d{\bf r}=0.
\label{mepp5}
\end{eqnarray}
Using Eqs. (\ref{b2}) and (\ref{mepp1}), we similarly express the rate of entropy production as
\begin{eqnarray}
\dot S=-\sum_{a} \int \frac{\nabla\omega_{a}}{\gamma_{a}\omega_{a}}\cdot {\bf J}_{a}\, d{\bf r}.
\label{mepp6}
\end{eqnarray}

The Maximum Entropy Production Principle (MEPP) consists in choosing the currents ${\bf J}_{a}$ which maximize the rate of entropy production $\dot S$ respecting the constraints $\dot E=0$, $\dot L=0$, $\dot {\bf P}={\bf 0}$, and $\sum_{a}{J_{a}^2}/({2\omega_{a}\gamma_{a}})\le C({\bf r},t)$. The last constraint expresses a bound (unknown) on the value of the diffusion currents. Convexity arguments justify that this bound is always reached so that the inequality can be replaced by an equality. The corresponding condition on first variations can be written at each time $t$ as
\begin{eqnarray}
\delta\dot S-\beta(t)\delta\dot E-\beta(t)\frac{\Omega(t)}{2}\delta\dot L+\beta(t){\bf U}(t)\cdot \delta\dot {\bf P}\nonumber\\
-\sum_{a}\frac{1}{D({\bf r},t)}\delta\left (\frac{J_{a}^2}{2\omega_{a}\gamma_{a}}\right )\, d{\bf r}=0,
\label{mepp7}
\end{eqnarray}
where $\beta(t)$, $\Omega(t)$, ${\bf U}(t)$, and $D({\bf r},t)$ are time-dependent Lagrange multipliers. The MEPP leads to optimal currents of the form
\begin{eqnarray}
{\bf J}_{a}=-D({\bf r},t) \left\lbrace \nabla\omega_{a}+\beta(t)\gamma_{a}\omega_{a}\left\lbrack\nabla\psi+\Omega(t){\bf r}-{\bf U}_{\perp}(t)\right\rbrack\right\rbrace.\nonumber\\
\label{mepp8}
\end{eqnarray}
We can also write
\begin{eqnarray}
{\bf J}_{a}=-D({\bf r},t) \left\lbrace \nabla\omega_{a}+\beta(t)\gamma_{a}\omega_{a}\left\lbrack {\bf u}-{\bf \Omega}(t)\times {\bf r}-{\bf U}(t)\right\rbrack_{\perp}\right\rbrace.\nonumber\\
\label{mepp8b}
\end{eqnarray}
Under this form, we see that the flow structures rotate with angular velocity
${\bf \Omega}(t)$ and/or translate with velocity ${\bf U}(t)$. The total current
is
\begin{eqnarray}
{\bf J}=-D({\bf r},t) \left\lbrace
\nabla\omega+\beta(t)\omega_{2}\left\lbrack\nabla\psi+\Omega(t){\bf r}-{\bf
U}_{\perp}(t)\right\rbrack\right\rbrace,\quad
\label{mepp9}
\end{eqnarray}
where $\omega_2=\sum_a\omega_a\gamma_a$ is the local microscopic enstrophy. Introducing the time-dependent relative stream function
\begin{eqnarray}
\psi_{eff}({\bf r},t)=\psi({\bf r},t)+\frac{\Omega(t)}{2}r^2-{\bf U}_{\perp}(t)\cdot {\bf r},
\label{mepp10}
\end{eqnarray}
we obtain the relaxation
equations
\begin{eqnarray}
\frac{\partial\omega_{a}}{\partial t}+{\bf u}\cdot \nabla\omega_{a}=\nabla\cdot \left\lbrace D({\bf r},t) \left\lbrack \nabla\omega_{a}+\beta(t)\gamma_{a}\omega_{a}\nabla\psi_{eff}\right\rbrack\right\rbrace.\nonumber\\
\label{mepp11}
\end{eqnarray}
The equation for the total vorticity is
\begin{eqnarray}
\frac{\partial\omega}{\partial t}+{\bf u}\cdot \nabla\omega=\nabla\cdot \left\lbrace D({\bf r},t) \left\lbrack \nabla\omega+\beta(t)\omega_{2}\nabla\psi_{eff}\right\rbrack\right\rbrace.
\label{mepp12}
\end{eqnarray}

The time evolution of the Lagrange multipliers $\beta(t)$, $\Omega(t)$, and ${\bf U}(t)$ is determined
by introducing Eq. (\ref{mepp9}) in the constraints (\ref{mepp3})-(\ref{mepp5}). This yields
\begin{equation}
\int D \nabla\psi\cdot \left\lbrace \nabla\omega+\beta(t)\omega_{2}\left\lbrack\nabla\psi+\Omega(t){\bf r}-{\bf U}_{\perp}(t)\right\rbrack\right\rbrace\, d{\bf r}=0,
\label{mepp13}
\end{equation}
\begin{equation}
\int D {\bf r}\cdot \left\lbrace \nabla\omega+\beta(t)\omega_{2}\left\lbrack\nabla\psi+\Omega(t){\bf r}-{\bf U}_{\perp}(t)\right\rbrack\right\rbrace\, d{\bf r}=0,
\label{mepp14}
\end{equation}
\begin{equation}
\int D  \left\lbrace \nabla\omega+\beta(t)\omega_{2}\left\lbrack\nabla\psi+\Omega(t){\bf r}-{\bf U}_{\perp}(t)\right\rbrack\right\rbrace\, d{\bf r}=0.
\label{mepp15}
\end{equation}
Equations (\ref{mepp13})-(\ref{mepp15}) form a system of linear equations that
determines the
evolution of $\beta(t)$, $\Omega(t)$, and ${\bf U}(t)$.
In a domain without special symmetry ($\Omega={\bf U}=0$), they reduce to
\begin{eqnarray}
\beta(t)=-\frac{\int D \nabla\psi\cdot\nabla\omega\, d{\bf r}}{\int D \omega_2 (\nabla\psi)^2\, d{\bf r}}.
\label{mepp16}
\end{eqnarray}
The evolution of the inverse temperature $\beta(t)$ is determined by
the conservation of energy.

The Boltzmann entropy (\ref{b2}) is the Lyapunov functional of the relaxation equations (\ref{mepp11}). It satisfies an $H$-theorem provided that $D({\bf r},t)\ge 0$. Indeed, using the expression (\ref{mepp8}) of the currents, the entropy production (\ref{mepp6}) can be rewritten as
\begin{eqnarray}
\dot S&=&\sum_{a} \int \frac{J_{a}^2}{D\gamma_{a}\omega_{a}}\, d{\bf r}\nonumber\\
&+&\beta(t)\int {\bf J}\cdot \left\lbrack\nabla\psi+\Omega(t){\bf r}-{\bf U}_{\perp}(t)\right\rbrack\, d{\bf r}.
\label{mepp17}
\end{eqnarray}
Using the conservation laws (\ref{mepp3})-(\ref{mepp5}), the second term is seen to vanish giving
\begin{eqnarray}
\dot S=\sum_{a} \int \frac{J_{a}^2}{D\gamma_{a}\omega_{a}}\, d{\bf r}\ge 0.
\label{mepp18}
\end{eqnarray}
A stationary solution of Eq. (\ref{mepp11}) satisfies $\dot S=0$ implying ${\bf J}_{a}={\bf 0}$ for each species. Using Eq. (\ref{mepp8}), we recover the mean field Boltzmann distribution (\ref{m3})
where $A_{a}$ appears as a constant of integration.

Because of the $H$-theorem, the relaxation equations converge, for $t\rightarrow
+\infty$, towards a mean field Boltzmann distribution that is a (local) maximum
of entropy at fixed energy, circulations, angular momentum, and linear
impulse.\footnote{The steady states of the relaxation equations are the critical
points (maxima, minima, saddle points) of the constrained entropy. It can be
shown \cite{pre} that a critical point of entropy is dynamically stable with
respect to the relaxation equations if,
and only if, it is a (local) maximum.  Minima are unstable for all perturbations
so they cannot be reached by the relaxation equations. Saddle points are
unstable only for certain perturbations so they can be reached if the system
does not spontaneously generate these dangerous perturbations. We note that
these properties are valid whatever the form of the diffusion coefficient
$D({\bf r},t)\ge 0$.} If several maxima exist for the same values of the
constraints, the selection depends on a notion of basin of attraction.

The relaxation equations adapted to the canonical ensemble can be obtained by
maximizing the rate of free energy production $\dot J$ with the constraint
$\sum_{a}{J_{a}^2}/({2\omega_{a}\gamma_{a}})\le C({\bf r},t)$. This leads to Eq.
(\ref{mepp11}) where $\beta$, $\Omega$ and ${\bf U}$ are fixed. These relaxation
equations increase the free energy (\ref{c2}) monotonically (canonical
$H$-theorem) and they relax towards a (local) maximum of free energy at fixed
circulations.

{\it Remark:} we stress that the relaxation equations (\ref{mepp11})
do {\it not} describe the real dynamics of
point vortices in the microcanonical ensemble. The kinetic theory of
point vortices developed in \cite{kinonsager} (and references therein) leads to very
different equations. Therefore, the relaxation equations (\ref{mepp11})
have probably no physical justification.\footnote{We note that
the relaxation equations (\ref{mepp11}) and the MEPP are closely related to the
linear thermodynamics of Onsager \cite{onsagerlinear}. The linear thermodynamics
of Onsager is expected to be valid close to equilibrium. However, the present
remark shows that this is not always the case since the rigorous kinetic theory
of 2D point vortices \cite{kinonsager} does {\it not} reduce to Onsager's linear
thermodynamics even close to equilibrium. The reason may be related to the fact
that Onsager's theory is valid in the canonical ensemble (see \cite{nfp} for
more details) while the kinetic theory of point vortices has to be developed in
the microcanonical ensemble. If we consider the kinetic theory of Brownian 
point vortices in the canonical ensemble \cite{pvbrownian}, we obtain
Fokker-Planck equations (see Sec. \ref{sec_bvz}) that are equivalent to
Onsager's linear thermodynamics \cite{nfp}.} However, since they relax towards
a (local) maximum of constrained entropy, they can be used as
a {\it numerical algorithm} to construct the statistical equilibrium
state of the multi-species point vortex gas for given values
of $E$, $\Gamma_a$, $L$ and ${\bf P}$. This confers them
some practical interest since it is not easy to directly solve the mean field
Boltzmann-Poisson equation (\ref{m8}) and be sure that the solution is stable (entropy maximum).

\subsection{The importance of the drift term}

The current ${\bf J}_a$ arising in the relaxation equations (\ref{mepp11}) is
the sum of two terms. A diffusion term ${\bf J}_{diff}=-D\nabla\omega_a$ and a
drift term ${\bf V}_a^{drift}=-D\beta\gamma_a\nabla\psi_{eff}$ that is
perpendicular to the relative mean field velocity ${\bf u}_{eff}=-{\bf z}\times
\nabla\psi_{eff}$. The drift coefficient $\mu_a=D\beta\gamma_a$ (mobility)
depends on the sign of the point vortices and is given by a sort of Einstein
relation. The diffusion term dissipates the energy ($\dot E_{diff}=\int {\bf
J}_{diff}\cdot\nabla\psi\, d{\bf r}=-\int D\nabla\omega\cdot\nabla\psi\, d{\bf
r}=-D\int\omega^2\, d{\bf r}\le 0$) while the drift term acts in the opposite
direction ($\dot E_{drift}=-\dot E_{diff}$) in order to conserve the energy
($\dot E=0$). On the other hand, the diffusion term increases the entropy ($\dot
S_{diff}=\sum_a \int D(\nabla\omega_a)^2/(\gamma_a\omega_a)\, d{\bf r}\ge 0$)
while the drift term usually decreases the entropy ($\dot S_{diff}\le 0$).
However, as a whole, the entropy increases ($\dot S\ge 0$). As emphasized in
previous papers \cite{csr,preR,pre,pvbrownian}, the diffusion term tends
to
disperse the vortices while the drift term is responsible for the clustering of
point vortices of the same sign at negative temperatures. It is therefore a
fundamental feature of two-dimensional
vortex dynamics. We
can also relate these results to the phenomenon of {\it inverse cascade} in 2D
turbulence. With only the diffusion term (which can be large since it represents
a turbulent viscosity), we have a direct cascade of energy towards smaller and
smaller scales. When the drift term is taken into account, we get an
inverse cascade of energy towards the large scales. By
contrast, the neg-entropy $-S$ (and more generally the generalized enstrophies
$\Gamma_n=\int\omega^n\, d{\bf r}$) always cascade towards the small scales and
are dissipated (see Sec. \ref{sec_ens}).

\subsection{The relaxation equations for two species of point vortices with opposite circulation}
\label{sec_twomepp}

If we consider a collection of $N/2$ vortices with circulation $+\gamma$ and
$N/2$ vortices
with circulation $-\gamma$, the  relaxation equations take
the form
\begin{equation}
\frac{\partial\omega_{\pm}}{\partial t}+{\bf u}\cdot \nabla\omega_{\pm}=\nabla\cdot \left\lbrace D \left\lbrack \nabla\omega_{\pm} \pm\beta(t)\gamma\omega_{\pm}\nabla\psi_{eff}\right\rbrack\right\rbrace,
\label{twomepp1}
\end{equation}
\begin{eqnarray}
-\Delta\psi=\omega=\omega_++\omega_-.
\label{twomepp2}
\end{eqnarray}
The evolution of the Lagrange multipliers is given by Eqs. (\ref{mepp13})-(\ref{mepp15}) with $\omega_2=(\omega_+-\omega_-)\gamma=n\gamma^2$. The equilibrium state is
\begin{eqnarray}
\omega_{\pm}=\pm A_{\pm}\gamma e^{\mp\beta\gamma\psi_{eff}}.
\label{twomepp3}
\end{eqnarray}
When $A_+=A_-$, this leads to the sinh-Poisson equation (\ref{sinh3}) but, as
indicated previously, this is not the general case.

\subsection{The relaxation equations for a single species of point vortices}
\label{sec_singlemepp}

If the point vortices have the same circulation $\gamma$ (single species
system), we find that the relaxation equation for the smooth vorticity field is
\begin{equation}
\frac{\partial\omega}{\partial t}+{\bf u}\cdot \nabla\omega=\nabla\cdot \left\lbrace D({\bf r},t) \left\lbrack \nabla\omega+\beta(t)\gamma\omega\nabla\psi_{eff}\right\rbrack\right\rbrace,
\label{mepp20}
\end{equation}
\begin{equation}
-\Delta\psi=\omega.
\label{mepp20xx}
\end{equation}
The time evolution of the Lagrange multipliers $\beta(t)$, $\Omega(t)$, and ${\bf U}(t)$ is given by \begin{equation}
\int D \nabla\psi\cdot \left\lbrace \nabla\omega+\beta(t)\gamma\omega\left\lbrack\nabla\psi+\Omega(t){\bf r}-{\bf U}_{\perp}(t)\right\rbrack\right\rbrace\, d{\bf r}=0,
\label{mepp21}
\end{equation}
\begin{equation}
\int D {\bf r}\cdot \left\lbrace \nabla\omega+\beta(t)\gamma\omega\left\lbrack\nabla\psi+\Omega(t){\bf r}-{\bf U}_{\perp}(t)\right\rbrack\right\rbrace\, d{\bf r}=0,
\label{mepp22}
\end{equation}
\begin{equation}
\int D  \left\lbrace \nabla\omega+\beta(t)\gamma\omega\left\lbrack\nabla\psi+\Omega(t){\bf r}-{\bf U}_{\perp}(t)\right\rbrack\right\rbrace\, d{\bf r}=0.
\label{mepp23}
\end{equation}
In a domain without special symmetry, they reduce to
\begin{eqnarray}
\beta(t)=-\frac{\int D \nabla\psi\cdot\nabla\omega\, d{\bf r}}{\int D \gamma\omega (\nabla\psi)^2\, d{\bf r}}.
\label{mepp24}
\end{eqnarray}
The conservation of energy determines the evolution of the inverse temperature $\beta(t)$. At equilibrium, we obtain the Boltzmann distribution
\begin{eqnarray}
\omega=A e^{-\beta\gamma\psi_{eff}}
\label{mepp25}
\end{eqnarray}
that maximizes the Boltzmann entropy
\begin{eqnarray}
S=-\int {\omega\over \gamma}\ln {\omega\over N\gamma} \ d{\bf r}
\label{mepp26}
\end{eqnarray}
at fixed energy, circulation, angular momentum, and linear impulse.

\subsection{Simplification of the constraints in an infinite domain}

We consider a single species of point vortices and we assume that the diffusion coefficient $D$  is constant in order to simplify the expressions of the constraints. In an infinite domain, we have the identities
\begin{eqnarray}
\label{mepp27}
\int\nabla{\omega}\cdot\nabla\psi\, d{\bf r}=\Gamma_2,\qquad Q\equiv \int{\omega}(\nabla\psi)^2\, d{\bf r},\nonumber\\
\int{\omega} {\bf r}\cdot \nabla\psi\, d{\bf r}={\cal V},\qquad
\int {\omega} \nabla\psi\, d{\bf r}={\bf 0}, \nonumber\\
\int {\bf r}\cdot \nabla {\omega} \, d{\bf r}=-2\Gamma, \qquad \int\nabla {\omega}\, d{\bf r}={\bf 0},
\end{eqnarray}
where
\begin{eqnarray}
\label{mepp28}
\Gamma_2=\int\omega^2\, d{\bf r}
\end{eqnarray}
is the macroscopic enstrophy and   the Virial ${\cal V}$ is given by Eq. (\ref{mfiv3}). With these expressions, the constraints  (\ref{mepp21})-(\ref{mepp23}) determining the evolution of the Lagrange multipliers $\beta(t)$, $\Omega(t)$ and ${\bf U}(t)$ become
\begin{eqnarray}
\label{mepp29}
\frac{\Gamma_2}{\beta\gamma}+Q+\Omega{\cal V}=0,
\end{eqnarray}
\begin{eqnarray}
\label{mepp30}
-\frac{2\Gamma}{\beta\gamma}+{\cal V}+\Omega L-{\bf U}\cdot {\bf P}=0,
\end{eqnarray}
\begin{eqnarray}
\label{mepp31}
\Omega {\bf P}-\Gamma {\bf U}={\bf 0}.
\end{eqnarray}
We can always impose ${\bf P}={\bf 0}$ by taking the origin at the center of
vorticity. Then, Eq. (\ref{mepp31}) implies ${\bf U}=(\Omega/\Gamma){\bf P}={\bf
0}$. The remaining
constraints reduce to
\begin{eqnarray}
\label{mepp32}
\frac{\Gamma_2}{\beta\gamma}+Q+\Omega{\cal V}=0,\qquad -\frac{2\Gamma}{\beta\gamma}+{\cal V}+\Omega L=0.
\end{eqnarray}
Solving these equations, we find that the Lagrange multipliers evolve according to
\begin{eqnarray}
\label{mepp33}
\beta(t)=\frac{\Gamma_2 L+2\Gamma{\cal V}}{\gamma({\cal V}^2-Q L)},\qquad \Omega(t)=-\frac{2\Gamma Q+{\cal V}\Gamma_2}{\Gamma_2 L+2\Gamma {\cal V}}.
\end{eqnarray}
Using the expression (\ref{mfiv3}) of the Virial, Eq. (\ref{mepp32}-b) may be
rewritten as
\begin{eqnarray}
\label{mepp34}
\Omega(t)=\frac{2\Gamma}{\gamma L}\left (\frac{1}{\beta(t)}-\frac{1}{\beta_c}\right ),
\end{eqnarray}
where $\beta_c=-{8\pi}/{(\gamma\Gamma)}$ is the critical inverse temperature (\ref{vmf6}). At equilibrium, we recover the Virial theorem of the point vortex gas (\ref{vmf4b}) in an unbounded domain. This relation remains valid out-of-equilibrium in the framework of the relaxation equations. We also note that $L=0$ does not imply $\Omega=0$.

In the canonical ensemble, the relaxation equations are given by Eqs. (\ref{mepp20})-(\ref{mepp20xx}) where $\beta$, $\Omega$, and ${\bf U}$ are fixed. We take ${\bf U}={\bf 0}$ for simplicity and we again assume that $D$ is constant. In that case, proceeding as in Sec. \ref{sec_virial}, we can establish the out-of-equilibrium Virial theorem
\begin{eqnarray}
\label{mepp34b}
\frac{1}{4\gamma D\beta}\dot L+\frac{1}{2}\Omega L=\frac{\Gamma}{\gamma}\left (\frac{1}{\beta}-\frac{1}{\beta_c}\right ).
\end{eqnarray}
At equilibrium, we recover the Virial theorem of the point vortex gas
(\ref{vmf4b}) in an unbounded domain. Additional results and discussion are
given in \cite{virialIJMPB}.

\subsection{Simplification of the constraints for an
axisymmetric flow in a disk}

In a disk, the constraints are given by Eqs. (\ref{mepp21})-(\ref{mepp22}) with ${\bf U}={\bf 0}$ since the linear impulse is not conserved. We again assume that $D$ is constant for simplicity. Solving these equations, we find that the Lagrange multipliers  $\beta(t)$ and $\Omega(t)$ evolve according to
\begin{eqnarray}
\label{ghd1}
\beta(t)=\frac{L\int\nabla\omega\cdot\nabla\psi\, d{\bf r}-{\cal V}\int {\bf r}\cdot\nabla\omega\, d{\bf r}}{\gamma {\cal V}^2-\gamma L\int \omega (\nabla\psi)^2\, d{\bf r}},
\end{eqnarray}
\begin{eqnarray}
\label{ghd2}
\Omega(t)=\frac{\int {\bf r}\cdot\nabla\omega\, d{\bf r}\int \omega (\nabla\psi)^2\, d{\bf r}-{\cal V}\int\nabla\omega\cdot\nabla\psi\, d{\bf r}}{\Gamma_2 L+2\Gamma {\cal V}}.
\end{eqnarray}
For an axisymmetric flow, we have the identities
\begin{eqnarray}
\label{mepp36}
\int\nabla {\omega}\cdot\nabla\psi\, d{\bf r}=\Gamma_2-\Gamma {\omega}_b,
\end{eqnarray}
\begin{eqnarray}
\label{mepp37}
\int {\bf r}\cdot \nabla {\omega} \, d{\bf r}=2\pi R^2 {\omega}_b-2\Gamma,
\end{eqnarray}
where $\omega_b=\omega(R,t)$ is the vorticity on the boundary of the disk. On the other hand, the Virial ${\cal V}$ is given by Eq. (\ref{mfiv3}). With these expressions, the constraints  (\ref{mepp21})-(\ref{mepp22}) reduce to
\begin{equation}
\label{mepp38}
\frac{\Gamma_2-\Gamma{\omega}_b}{\beta\gamma}+Q+\Omega{\cal V}=0,\quad \frac{2\pi R^2\omega_b-2\Gamma}{\beta\gamma}+{\cal V}+\Omega L=0,\nonumber\\
\end{equation}
and the Lagrange multipliers $\beta(t)$ and $\Omega(t)$ evolve according to
\begin{eqnarray}
\label{mepp39}
\beta(t)=\frac{(\Gamma_2-\Gamma {\omega}_b) L+2(\Gamma-\pi R^2 {\omega}_b){\cal V}}{\gamma({\cal V}^2-QL)},
\end{eqnarray}
\begin{eqnarray}
\label{mepp40}
\Omega(t)=-\frac{2(\Gamma-\pi R^2 {\omega}_b) Q+{\cal V}(\Gamma_2-\Gamma {\omega}_b)}{(\Gamma_2-\Gamma {\omega}_b) L+2(\Gamma-\pi R^2 {\omega}_b) {\cal V}}.
\end{eqnarray}
Eq. (\ref{mepp38}-b) may be rewritten as
\begin{eqnarray}
\label{mepp41}
\frac{{\omega}_b(t)}{\beta(t)\gamma}\pi R^2=N\left (\frac{1}{\beta(t)}-\frac{1}{\beta_c}\right )-\frac{1}{2}\Omega(t) L,
\end{eqnarray}
where $\beta_c$ is the critical inverse temperature (\ref{vmf6}). At
equilibrium, we recover the Virial theorem (\ref{vmf4}) of the axisymmetric point
vortex gas in a disk. This relation remains valid
out-of-equilibrium in the framework of the relaxation equations.

In the canonical ensemble, the relaxation equations are given by Eqs. (\ref{mepp20})-(\ref{mepp20xx}) where $\beta$ and $\Omega$ are fixed and ${\bf U}={\bf 0}$. In that case, proceeding as in Sec. \ref{sec_virial}, we can establish the out-of-equilibrium Virial theorem
\begin{eqnarray}
\label{vca1}
\frac{1}{2\gamma D\beta}\dot L=\frac{2\Gamma}{\gamma\beta}-{\cal V}-\Omega L-2 P V.
\end{eqnarray}
For an axisymmetric flow, using the expression  (\ref{mfiv3}) of the Virial, we get
\begin{eqnarray}
\label{vca2}
\frac{1}{4\gamma D\beta}\dot L=\frac{\Gamma}{\gamma}\left (\frac{1}{\beta}-\frac{1}{\beta_c}\right )-\frac{1}{2}\Omega L-\frac{\omega_b(t)}{\gamma\beta}\pi R^2.
\end{eqnarray}
At equilibrium, we recover the Virial theorem (\ref{vmf4}) of the axisymmetric point
vortex gas in a disk. Additional results and discussion are given in \cite{virialIJMPB}.

\subsection{Connection with other relaxation equations}
\label{sec_conn}

In this section, we mention some connections (but also stress important differences) between the relaxation equations derived in the previous sections and  other  equations of the same kind. To simplify the discussion, we take $\Omega=0$ and ${\bf U}={\bf 0}$.

\subsubsection{Relaxation of a test vortex in a thermal bath}

The relaxation equation (\ref{mepp20}) shares some analogies with the Fokker-Planck equation
\begin{equation}
\frac{\partial P}{\partial t}+{\bf u}_{bath}\cdot \nabla P=\nabla\cdot
\left\lbrace D({\bf r}) \left\lbrack \nabla P+\beta\gamma
P\nabla\psi_{bath}\right\rbrack\right\rbrace
\label{conn1}
\end{equation}
derived in the kinetic theory of point vortices in the thermal bath approximation \cite{kinonsager}.  The Fokker-Planck equation (\ref{conn1}) describes the evolution of the probability density $P({\bf r},t)$ of the position ${\bf r}$ of a test vortex evolving in a ``bath'' of field vortices at statistical equilibrium. The stream function $\psi_{bath}({\bf r})$ in Eq. (\ref{conn1}) is {\it fixed}. It is determined by the equilibrium distribution $\omega_{bath}({\bf r})=Ae^{-\beta\gamma\psi_{bath}({\bf r})}$ of the field vortices forming the bath.  The probability current in Eq. (\ref{conn1})  is the sum of two terms. A term $D\nabla P$ corresponding to a pure diffusion with a diffusion coefficient $D$ (its expression is given in \cite{kinonsager}) and an additional term $D\beta\gamma\nabla\psi_{bath}$ that can be interpreted as a {\it drift} velocity.\footnote{There exist numerous analogies between the kinetic theory of 2D vortices and the kinetic theory of stellar systems \cite{houches}. In these analogies, the drift of a point vortex \cite{preR} is the counterpart of the dynamical friction experienced by a star \cite{chandrafriction}.} The drift is perpendicular to the mean velocity ${\bf u}_{bath}=-{\bf z}\times\nabla\psi_{bath}$. The drift coefficient  $\mu=D\beta\gamma$ (mobility) is given by a sort of Einstein relation. At equilibrium, we get the Boltzmann distribution
\begin{eqnarray}
P_{eq}({\bf r})=A'e^{-\beta\gamma\psi_{bath}({\bf r})}.
\label{conn1b}
\end{eqnarray}
Therefore, the distribution of the test vortex relaxes towards the distribution of the field vortices (bath). As a result, the Fokker-Planck equation (\ref{conn1}) describes a process of  thermalization.

Despite some obvious analogies, the relaxation equations  (\ref{mepp20}) and
(\ref{conn1}) are very different (physically and mathematically). In  Eq.
(\ref{mepp20}) the stream function $\psi({\bf r},t)$ is determined
self-consistently by the vorticity $\omega({\bf r},t)$ using the Poisson
equation  (\ref{mepp20xx}). As a result, it relaxes towards the Boltzmann
distribution (\ref{mepp25}) coupled self-consistently to  the Poisson equation
(\ref{mepp20xx}). By contrast, in Eqs. (\ref{conn1}) and  (\ref{conn1b}) the
stream function $\psi_{bath}({\bf r})$ is given. 

\subsubsection{Brownian vortices}
\label{sec_bvz}

In the canonical ensemble (fixed  $\beta$), the relaxation  equations (\ref{mepp11}) coincide with the Fokker-Planck equations
\begin{eqnarray}
\frac{\partial\omega_{a}}{\partial t}+{\bf u}\cdot \nabla\omega_{a}=\nabla\cdot \left\lbrace D\left\lbrack \nabla\omega_{a}+\beta\gamma_{a}\omega_{a}\nabla\psi\right\rbrack\right\rbrace,
\label{mepp11ex}
\end{eqnarray}
\begin{eqnarray}
-\Delta\psi=\omega=\sum_{a}\omega_{a}
\label{mf1ex}
\end{eqnarray}
describing a multi-species system of ``Brownian point vortices'' in the mean field approximation \cite{pvbrownian}. At equilibrium, we get the Boltzmann-Poisson equation
\begin{eqnarray}
-\Delta\psi=\sum_a A_a \gamma_a e^{-\beta \gamma_a \psi}.
\label{conn3bex}
\end{eqnarray}
The temperature may be positive or negative. At negative temperatures, like-sign Brownian vortices attract each other and at positive temperatures like-sign Brownian vortices repel each other (in this model, the temperature is the relevant control parameter).

\subsubsection{Smoluchowski-Poisson system and Keller-Segel model}

The relaxation equations (\ref{mepp11ex})-(\ref{mf1ex}) with $\gamma_a>0$ and $\beta<0$ are similar to the mean field Fokker-Planck equations
\begin{eqnarray}
\frac{\partial\rho_{a}}{\partial t}=\nabla\cdot \left\lbrack D \left ( \nabla\rho_{a}+\beta m_{a}\rho_{a}\nabla\Phi\right )\right\rbrack,
\label{conn2}
\end{eqnarray}
\begin{eqnarray}
\Delta\Phi=S_d G\rho=S_d G\sum_a\rho_a=S_d G\sum_a n_a m_a
\label{conn3}
\end{eqnarray}
describing  a multi-species system of self-gravitating Brownian particles 
in the strong friction limit \cite{sopik}. These equations form the so-called
Smoluchowski-Poisson system. Here, $m_a$ is the mass of particles of species
$a$,  $\rho_a$ is the density of mass of species $a$, $\Phi$ is the
gravitational potential, and $\beta=1/(k_B T)>0$ is the inverse temperature
($S_d$ is the surface of a unit sphere in $d$ dimensions). At equilibrium, we
get the Boltzmann-Poisson equation
\begin{eqnarray}
\Delta\Phi=S_d G\sum_a A_a m_a e^{-\beta m_a \Phi}.
\label{conn3b}
\end{eqnarray}
Of course, the mass of the particles is always positive and the gravitational
force is attractive. 
As discussed in
\cite{sopik}, the relaxation equations (\ref{conn2})-(\ref{conn3}) also provide
a  multi-species Keller-Segel \cite{ks} model of
chemotaxis in biology. With the notations of biology, it writes
\begin{eqnarray}
\frac{\partial\rho_{a}}{\partial t}=D_a
\Delta\rho_{a}-\chi_a \nabla(\rho_a \nabla c),
\label{bioconn2}
\end{eqnarray}
\begin{eqnarray}
\Delta c=-\lambda \rho=-\lambda \sum_a\rho_a,
\label{bioconn3}
\end{eqnarray}
where $\rho_a$ is the
density of particles of species $a$ and  $c$  is the concentration of the
secreted chemical (pheromone). The particles diffuse with a
diffusion coefficient $D_a$ and they also move in the direction of
the gradient of the secreted chemical (chemotactic drift). The chemotactic
sensitivity $\chi_a$ is a measure of the strength of the influence of the
chemical gradient on the flow of particles. Finally, the coefficient $\lambda$
measures the quantity of chemical produced by the particles. At equilibrium,
we get the Boltzmann-Poisson equation
\begin{eqnarray}
\Delta c=-\lambda \sum_a A_a  e^{\frac{\chi_a}{D_a}c}
\label{bioconn3b}
\end{eqnarray}
that is equivalent to Eq. (\ref{conn3b}) up to a change of notations. In
this analogy, the concentration $-c$ of the chemical plays the role of the
gravitational potential $\Phi$.

\subsubsection{Debye-H\"uckel equations}

The relaxation equations (\ref{mepp11ex})-(\ref{mf1ex}) with  $\beta>0$ are similar to the mean field Fokker-Planck equations
\begin{eqnarray}
\frac{\partial\rho_{a}}{\partial t}=\nabla\cdot \left\lbrack D \left (\nabla\rho_{a}+\beta e_{a}\rho_{a}\nabla\Phi\right )\right\rbrack,
\label{conn4}
\end{eqnarray}
\begin{eqnarray}
\Delta\Phi=-S_d \rho=-S_d \sum_a\rho_a=-S_d \sum_a n_a e_a
\label{conn5}
\end{eqnarray}
derived by Debye and H\"uckel \cite{dh2} in their theory of
electrolytes.\footnote{These equations were previously derived by Nernst
\cite{nernst1,nernst2} and Planck \cite{planck} so they are also called
the Nernst-Planck equations.} Here, $e_a$ is
the charge of particles of species $a$, $\rho_a$ is the density of charge of
species $a$, $\Phi$ is the electric potential, and $\beta=1/(k_B T)>0$ is the
inverse temperature. At equilibrium, we get the Boltzmann-Poisson equation
\begin{eqnarray}
\Delta\Phi=-S_d\sum_a A_a e_a e^{-\beta e_a \Phi}.
\label{conn6}
\end{eqnarray}
Particles of opposite sign attract each other while particles of the same sign
repel each other. If we consider a globally neutral system of charges with
$N/2$ charges $+e$ and $N/2$ charges $-e$, and assume $A_+=A_-=A$, we find that
the Boltzmann-Poisson
equation (\ref{conn6}) takes the form $\Delta\Phi=2S_d A e \sinh(\beta e\Phi)$ 
similar to the sinh-Poisson equation (\ref{sinh3}) but with positive
temperature. This sinh-Poisson equation explicitly appears in the paper of Debye
and H\"uckel \cite{dh1}.

In conclusion, the relaxation equations (\ref{mepp11}) for the point vortex gas at negative temperatures share some analogies with the Smoluchowski-Poisson system and with the Keller-Segel model, while the relaxation equations (\ref{mepp11}) for the point vortex gas at positive temperatures share some analogies with  the Debye-H\"uckel equations. However, we note that in Eqs. (\ref{mepp11}) the energy is conserved  thanks to a time dependent inverse temperature $\beta(t)$ (microcanonical ensemble) while the inverse temperature $\beta$ is fixed in the other equations (canonical ensemble).

\section{The strong mixing limit}
\label{sec_sm}

In this section, we consider the case where $\beta\gamma_a\psi\ll 1$ so that the
equations of the statistical theory can be expanded as a function of this small
parameter. This corresponds to a  limit of strong mixing ($\beta$ small) or low
energy ($\psi$ small) where the vorticity is almost uniform. This is also
similar to the Debye-H\"uckel approximation in plasma physics, except that in
the present context the temperature may be positive or negative. This expansion
was introduced by Chavanis and Sommeria \cite{jfm1} in the context of the MRS
statistical theory of continuous vorticity flows. We adapt it here to the case
of  point vortices and discuss the similarities and the differences with the
case of continuous vorticity flows. In this section, we present
the main results of the expansion. The details of the derivation, and some
complements, are given in Appendices \ref{sec_smpv} and \ref{sec_smcv}.

\subsection{Statistical equilibrium state}
\label{sec_dh}

We consider a multi-species system of point vortices in a bounded domain. We introduce appropriate length and time scales such that the area of the domain is $V=1$ and the microscopic enstrophy is $\Gamma_2^{m}=\sum_a \Gamma_a \gamma_a=\sum_a N_a \gamma_a^2=1$.  At statistical equilibrium, the relation between the vorticity and the stream function is given by Eq. (\ref{m3}). Assuming $\beta\gamma_{a}\psi\ll 1$, we obtain to second order
\begin{eqnarray}
\label{dh1}
\omega_{a}=A_a \gamma_a \left (1-\beta\gamma_a\psi+\frac{1}{2}\beta^2\gamma_a^2\psi^2+...\right ).
\end{eqnarray}
Integrating over the domain, we get
\begin{eqnarray}
\label{dh2}
\Gamma_{a}=A_a \gamma_a \left (1-\beta\gamma_a\langle\psi\rangle+\frac{1}{2}\beta^2\gamma_a^2\langle \psi^2\rangle+...\right ).
\end{eqnarray}
This equation can be reversed to give
\begin{equation}
\label{dh3}
A_{a}=N_a \left (1+\beta\gamma_a\langle\psi\rangle-\frac{1}{2}\beta^2\gamma_a^2\langle \psi^2\rangle+\beta^2\gamma_a^2\langle\psi\rangle^2+...\right ).
\end{equation}
At first order, the relation between the vorticity and the stream function is linear. Combining Eqs. (\ref{dh1}) and (\ref{dh3}), we get
\begin{eqnarray}
\label{dh4}
\omega_{a}=\Gamma_a -\beta\gamma_a\Gamma_a(\psi-\langle\psi\rangle).
\end{eqnarray}
The total vorticity is
\begin{eqnarray}
\label{dh5}
\omega=\Gamma -\beta (\psi-\langle\psi\rangle).
\end{eqnarray}
Substituting this relation in the Poisson equation (\ref{mf1}), we obtain the Helmholtz-type mean field equation
\begin{eqnarray}
\label{dh5b}
-\Delta\psi=\Gamma -\beta (\psi-\langle\psi\rangle).
\end{eqnarray}
The inverse temperature $\beta$ is determined by the energy constraint.
Using Eqs. (\ref{mf2}-a) and (\ref{dh5}), we get
\begin{eqnarray}
\label{dh6}
E=\frac{1}{2}\Gamma\langle\psi\rangle-\frac{1}{2}\beta (\langle\psi^2\rangle-\langle\psi\rangle^2).
\end{eqnarray}
The mean field equation (\ref{dh5b}) with the energy constraint (\ref{dh6}) may have several solutions with the same values of $\Gamma$ and $E$. To compare these solutions, we need to determine the entropy at second order. Substituting Eq. (\ref{m3}) in Eq. (\ref{b2}), we obtain
\begin{eqnarray}
\label{dh7}
S=2\beta E+\sum_a\alpha_a\Gamma_a+\sum_a N_a\ln N_a+N\nonumber\\
=2\beta E-\sum_a \frac{\Gamma_a}{\gamma_a}\ln\left (\frac{A_a}{N_a}\right ).
\end{eqnarray}
Using the expression of $A_a$ at second order given by Eq. (\ref{dh3}), we find that
\begin{eqnarray}
\label{dh8}
S=2\beta E-\beta\Gamma\langle\psi\rangle+\frac{1}{2}\beta^2(\langle\psi^2\rangle-\langle\psi\rangle^2).
\end{eqnarray}
Using Eq. (\ref{dh6}) to simplify the expression, we get
\begin{eqnarray}
\label{dh9}
S=-\frac{1}{2}\beta^2(\langle\psi^2\rangle-\langle\psi\rangle^2)=\beta
E-\frac{1}{2}\beta\Gamma\langle\psi\rangle.
\end{eqnarray}
On the other hand, using Eq. (\ref{dh5}),  the macroscopic enstrophy
(\ref{mepp28}) is given at second order by\footnote{In order to establish the
expression of the enstrophy at second order, we need to write the $\omega-\psi$
relationship at second order. However, it is shown in Appendix \ref{sec_smpv}
that the term of second order in the $\omega-\psi$ relationship does not
contribute to the enstrophy at second order leaving Eq. (\ref{dh11}) unchanged.}
\begin{eqnarray}
\label{dh11}
\Gamma_2=\Gamma^2+\beta^2(\langle\psi^2\rangle-\langle\psi\rangle^2)=\Gamma^2+
\beta\Gamma\langle\psi\rangle-2\beta E.\quad
\end{eqnarray}
Comparing Eqs. (\ref{dh9}) and (\ref{dh11}), we obtain the following relation between the entropy and the macroscopic enstrophy
\begin{eqnarray}
\label{dh12}
S=-\frac{1}{2}(\Gamma_2-\Gamma^2).
\end{eqnarray}
Therefore, in the strong mixing limit, a maximum of entropy is equivalent to a minimum of enstrophy.
At leading order, these results are exactly the same  as those obtained by Chavanis and Sommeria \cite{jfm1} for continuous vorticity flows in the MRS theory.

The expansion can be pushed to higher orders by a similar method (see Appendix \ref{sec_smpv}). We then obtain
\begin{eqnarray}
\label{dh13}
\omega=-\Delta\psi=B_0+B_1\beta\psi+B_2\beta^2\psi^2+B_3\beta^3\psi^3
\end{eqnarray}
with
\begin{eqnarray}
\label{dh14}
B_2=\frac{1}{2}\Gamma_3^m+\frac{1}{2}\Gamma_4^m\beta\langle\psi\rangle
\end{eqnarray}
and
\begin{eqnarray}
\label{dh15}
B_3=-\frac{1}{6}\Gamma_4^m,
\end{eqnarray}
where $\Gamma_n^m=\sum_a N_a\gamma_a^n$ denotes the microscopic moment of order $n$ of the vorticity. The coefficients $B_0$ and $B_1$ can be obtained from the general results given in Appendix \ref{sec_smpv}. In the case of a symmetric distribution of point vortices with positive and negative circulations, we have $\Gamma_{2n+1}^m=0$ and Eq. (\ref{dh13}) reduces to
\begin{equation}
\label{dh16}
\omega=B_0+B_1\beta\psi+\frac{1}{2}\Gamma_4^m\beta^3\langle\psi\rangle\psi^2
-\frac{1}{6}\Gamma_4^m \beta^3\psi^3.
\end{equation}
If in addition $\langle\psi^{2n+1}\rangle=0$, the even terms in $\psi$
disappear, and we obtain (using the results of Appendix \ref{sec_smpv}):
\begin{eqnarray}
\label{dh17}
\omega=-\left (\Gamma_2^m-\frac{1}{2}\Gamma_4^m\beta^2\langle\psi^2\rangle\right )\beta\psi-\frac{1}{6}\Gamma_4^m \beta^3\psi^3.
\end{eqnarray}
These expressions differ from those derived by Chavanis and Sommeria \cite{jfm1}
in the framework of the MRS theory (see Appendix A of \cite{jfm1} and Appendix
\ref{sec_smcv}). In particular,  the prefactor
$\Gamma_4^m$ of the cubic term in Eqs. (\ref{dh16}) and (\ref{dh17}) is
replaced, for continuous
flows, by 
$\Gamma_4^{f.g.}-3(\Gamma_2^{f.g.})^2$ where
$\Gamma_n^{f.g.}=\int\overline{\omega^n}\, d{\bf r}$ are the microscopic moments
of the vorticity (see \cite{jfm1} and Eq. (\ref{smcv21})). It is then found that
the Kurtosis ${\rm
Ku}=\Gamma_4^{f.g.}/(\Gamma_2^{f.g.})^2$ controls the deviation to the linear
$\omega-\psi$ relationship \cite{jfm1}. Consider for simplicity the case
$\langle\psi^{2n+1}\rangle=0$. For continuous vorticity flows, the $\omega-\psi$
relationship behaves like a tanh function for ${\rm Ku}<3$ and like a sinh
function for ${\rm Ku}>3$.
In the case of point vortices, there is no such transition and the 
$\omega-\psi$ relationship always behaves like a sinh function.

It is also interesting to consider the case of a system with $N/2$ point vortices of circulation $\gamma$ and $N/2$ point vortices of circulation $-\gamma$. In that case $\Gamma_2^m=N\gamma^2$ and $\Gamma_4^m=N\gamma^4$, and Eq. (\ref{dh16}) becomes
\begin{equation}
\label{dh18}
\omega=B_0+B_1\beta\psi+\frac{1}{2}N\gamma^4\beta^3\langle\psi\rangle\psi^2
-\frac{1}{6}N\gamma^4 \beta^3\psi^3.
\end{equation}
When $\langle \psi^{2n+1}\rangle=0$, we get from Eq. (\ref{dh17}):
\begin{eqnarray}
\label{dh17b}
\omega=-N\gamma^2\left (1-\frac{1}{2}\gamma^2\beta^2\langle\psi^2\rangle\right )\beta\psi-\frac{1}{6}N\gamma^4 \beta^3\psi^3.
\end{eqnarray}
This expansion can be directly obtained from the
sinh-Poisson equation (\ref{sinh3}). However, Eq. (\ref{dh18}) shows the
deviations to the sinh-Poisson equation induced by a non-vanishing
value of $\langle\psi\rangle$. Actually, Eq. (\ref{dh5b}) shows that the effect
of $\langle\psi\rangle$ is already present
at the linear order (see Sec. \ref{sec_geo}).

\subsection{Connection with other approaches}
\label{sec_co}

\subsubsection{The Debye-H\"uckel approximation}
\label{sec_dha}

At leading order, the expansion of the equations of the statistical theory as a function of $\beta\gamma_a\psi\ll 1$ leads to an equation of the form
\begin{eqnarray}
\label{dh19}
\Delta\psi-\frac{\beta \Gamma_2^m}{V}\psi=-\frac{\Gamma}{V}-\frac{\beta \Gamma_2^m}{V}\langle\psi\rangle,
\end{eqnarray}
where we have restored the domain area $V$ and the microscopic enstrophy
$\Gamma_2^m$. This expansion is similar to the Debye-H\"uckel approximation in
their theory of electrolytes \cite{dh1} except that in the present situation the
temperature may be positive or negative. At positive temperatures, and for a
neutral system, like-sign vortices ``repel'' each other and opposite-sign
vortices ``attract'' each other. A point vortex of a given sign is surrounded by
a cloud of point vortices of the opposite sign so the interaction between two
vortices is shielded on a length scale $\lambda_D=(\beta\sum_a
n_a\gamma_a^2)^{-1/2}$ equivalent to the Debye length.\footnote{In their
original paper, Debye and H\"uckel \cite{dh1} explicitly write the multi-species
Boltzmann-Poisson equation (\ref{m8}) and the sinh-Poisson equation
(\ref{sinh3}) with $\beta>0$ and $A_a=n_a$. They linearize these equations at
high temperature leading to Eq. (\ref{dh19}) with the right hand side equal to
zero. This is how the Debye screening length appears in their theory. A more
precise justification of the Debye length is given in Appendix \ref{sec_debye}.}
As a result of this shielding, the system is spatially homogeneous. At negative
temperatures, like-sign vortices ``attract'' each other and opposite-sign
vortices ``repel'' each other. They form large scale structures at the system
length scale in order to maximize their entropy (see \cite{jfm1} and Sec.
\ref{sec_geo}). This corresponds to the phenomenon of ``condensation''
discovered by Kraichnan \cite{kraichnan} with his statistical theory of 2D
turbulence in spectral space. This is also a direct consequence of the general
arguments of statistical mechanics given by Onsager \cite{onsager}.

\subsubsection{The minimum enstrophy principle}
\label{sec_ens}

The strong mixing (or low energy) limit allows us to make a connection between the rigorous maximum entropy principle of statistical mechanics \cite{jm} and the phenomenological minimum enstrophy principle  \cite{leith} introduced for a slightly viscous flow.\footnote{For slightly viscous flows, the enstrophy cascades towards the small scales where it is dissipated while the energy remains blocked in the largest scales (inverse cascade) and is well-conserved. This {\it selective decay} has motivated the minimum enstrophy principle where it is argued that the flow tends to minimize enstrophy at fixed energy (and circulation).} Indeed, in this limit, the statistical theory leads to a linear $\omega-\psi$ relationship (\ref{dh5}), like when we minimize the enstrophy at fixed energy and circulation, writing the first variations as
\begin{eqnarray}
\label{dh20}
\delta\Gamma_2+2\beta\delta E+\alpha\delta\Gamma=0,
\end{eqnarray}
where $\beta$ and $\alpha$ are Lagrange multipliers. Furthermore, Eq. (\ref{dh12}) shows that a maximum entropy state corresponds to a minimum enstrophy state. In the case of point vortices there is no viscosity but the macroscopic enstrophy is not conserved and can decrease. The same connection between maximum entropy states and minimum enstrophy states is found for continuous vorticity flows in the limit of strong mixing \cite{jfm1}. In the inviscid MRS theory, the microscopic enstrophy is conserved but the macroscopic enstrophy calculated with the coarse-grained vorticity $\overline{\omega}({\bf r})$ is not conserved and can decrease.

\subsubsection{The energy-enstrophy ensemble}
\label{sec_eee}

In the context of the MRS theory, a linear  $\omega-\psi$ relationship can be obtained by maximizing the MRS entropy at fixed circulation, energy, and microscopic enstrophy (neglecting the conservation of the higher order moments of the vorticity). This approach has been developed in \cite{naso}. In that case, the vorticity distribution is Gaussian and the mean flow is characterized by a linear  $\omega-\psi$ relationship. Furthermore, it can be proven that a maximum entropy state is equivalent to a minimum enstrophy state (see Appendix A of \cite{naso}). This approach may be seen as a formulation of the statistical mechanics of  Kraichnan  \cite{kraichnan} based on the ``energy-enstrophy'' ensemble in terms of the more modern MRS theory.

\subsection{Geometry induced phase transitions}
\label{sec_geo}

The linear mean field equation (\ref{dh5b}) with the quadratic energy constraint (\ref{dh6}) has been solved in \cite{jfm1} by expanding the stream function on the eigenmodes of the Laplacian. For small values of the control parameter $\Lambda=\Gamma/\sqrt{2E}$, one finds several solutions. In that case, we have to select the solution with the highest entropy. The general problem has been solved in \cite{jfm1}. If we consider the particular case $\Gamma=0$, one obtains two types of solution: a ``dipole'' ($\langle\psi\rangle=0$) which is an eigenfunction of the Laplacian and a ``monopole'' ($\langle\psi\rangle\neq 0$). Note that if we use the sinh-Poisson equation (\ref{sinh3}), we only get the dipole solution ($\langle\psi\rangle=0$) and the monopole solution ($\langle\psi\rangle\neq 0$) is forgotten. This shows the limitation of this equation. Chavanis and Sommeria \cite{jfm1} have found that the selection of the maximum entropy state depends on the geometry of the domain. For example, in a rectangular domain, there is a critical aspect ratio $\tau_c=1.12$. For $\tau<\tau_c$ the maximum entropy state is the ``monopole'' while for  $\tau>\tau_c$ the maximum entropy state is the ``dipole'' (it can be shown furthermore that the state with the lower value of the entropy is a saddle point). This gives rise to geometry induced phase transition between ``monopoles'' and ``dipoles''.

In a more recent paper, Taylor {\it et al.} \cite{taylor} have remarked that the solutions with $\langle\psi\rangle=0$ have zero angular momentum while the solutions with $\langle\psi\rangle\neq 0$ have nonzero angular momentum, even though their circulation is zero. Therefore, the ``dipoles'' are ``zero-momentum'' states and the ``monopoles'' are ``nonzero-momentum'' states or ``spin-up'' states. As a result, as discussed by Taylor {\it et al.} \cite{taylor}, the geometry induced phase transition between ``monopoles'' and ``dipoles'' found by Chavanis and Sommeria \cite{jfm1} can explain the ``spin-up'' phenomenon observed experimentally by Clercx {\it et al.} \cite{clercx}.

The methodology of Chavanis and Sommeria \cite{jfm1} has been generalized
recently to several systems: axisymmetric flows \cite{axi1,axi2}, Fofonoff flows
in oceanic basins \cite{fofonoff1,fofonoff2}, and barotropic flows on a sphere
\cite{sphere1,sphere2}. On the other hand, Venaille and Bouchet
\cite{vb} have shown that the phase transitions between ``monopoles'' and
``dipoles'' found by Chavanis and Sommeria \cite{jfm1} could be interpreted in terms of a ``bicritical point'' and  a ``second order
azeotropy''.

\subsection{Relaxation equations}
\label{sec_dhr}

It is interesting to see how the relaxation equations (\ref{mepp11}) of the point vortices can be simplified in the limit of strong mixing (or low energy). At leading order, it suffices to make the approximation $\omega_a\simeq \Gamma_a$ in the drift term. This yields
\begin{equation}
\frac{\partial\omega_{a}}{\partial t}+{\bf u}\cdot \nabla\omega_{a}=\nabla\cdot \left\lbrace D({\bf r},t) \left\lbrack \nabla\omega_{a}+\beta(t)\gamma_{a}\Gamma_{a}\nabla\psi\right\rbrack\right\rbrace.
\label{meppd1}
\end{equation}
The equation for the total vorticity is therefore
\begin{eqnarray}
\frac{\partial\omega}{\partial t}+{\bf u}\cdot \nabla\omega=\nabla\cdot \left\lbrace D({\bf r},t) \left\lbrack \nabla\omega+\beta(t)\nabla\psi\right\rbrack\right\rbrace
\label{meppd2}
\end{eqnarray}
corresponding to the vorticity current
\begin{eqnarray}
{\bf J}=-D({\bf r},t) \left\lbrack \nabla\omega+\beta(t)\nabla\psi\right\rbrack.
\label{meppd3}
\end{eqnarray}
We note that this equation is {\it closed} contrary to Eq. (\ref{mepp12}) when
we are
not in the strong mixing limit.  The conservation
of energy determines the evolution of the inverse temperature $\beta(t)$.
Substituting Eq. (\ref{meppd3}) in Eq. (\ref{mepp3}), we get
\begin{eqnarray}
\beta(t)=-\frac{\int D \nabla\psi\cdot\nabla\omega\, d{\bf r}}{\int D  (\nabla\psi)^2\, d{\bf r}}.
\label{meppd4}
\end{eqnarray}

Finally, we can show that the macroscopic enstrophy (\ref{mepp28}) is the Lyapunov functional of the relaxation equation (\ref{meppd2}). Indeed, it satisfies an $H$-theorem provided that $D\ge 0$. Using Eq. (\ref{mepp2}), the rate of enstrophy dissipation is given by
\begin{eqnarray}
\dot\Gamma_2=2\int {\bf J}\cdot \nabla\omega\, d{\bf r}.
\label{meppd5}
\end{eqnarray}
Using the expression (\ref{meppd3}) of the current, it can be rewritten as
\begin{eqnarray}
\dot \Gamma_2=-2 \int \frac{J^2}{D}\, d{\bf r}-2\beta(t)\int {\bf J}\cdot \nabla\psi\, d{\bf r}.
\label{meppd6}
\end{eqnarray}
Using the conservation of energy (\ref{mepp3}), the second term is seen to vanish giving
\begin{eqnarray}
\dot \Gamma_2=-2 \int \frac{J^2}{D}\, d{\bf r}\le 0.
\label{meppd7}
\end{eqnarray}
A stationary solution of Eq. (\ref{meppd2}) satisfies $\dot \Gamma_2=0$ implying ${\bf J}={\bf 0}$. Using Eq. (\ref{meppd3}), we recover the minimum enstrophy state (\ref{dh5}). Because of the $H$-theorem, the relaxation equation (\ref{meppd2}) converges, for $t\rightarrow +\infty$, towards a (local) minimum of enstrophy at fixed energy and  circulation. If several minima exist for the same values of the constraints, the selection depends on a notion of basin of attraction.

{\it Remark:} The relaxation equation (\ref{meppd2}) with (\ref{meppd4}) can also be obtained by maximizing the rate of dissipation of enstrophy at fixed circulation and energy \cite{gen}. This kind of relaxation equations has been numerically solved in \cite{naso,nasoeht} in order to illustrate the phase transitions discussed in Sec. \ref{sec_geo}.

\section{Conclusion}

We have complemented the literature on the statistical
mechanics of point vortices in two-dimensional hydrodynamics. We have derived
the Boltzmann-Poisson equation for a multi-species gas of point vortices by
using the maximum entropy principle that we have justified from the theory of
large deviations. We have also derived relaxation equations towards the maximum
entropy state by using a maximum entropy production principle. These equations
do not describe the true dynamics of the point vortex gas (this requires to
develop a more complicated kinetic theory of point vortices \cite{kinonsager})
but they can be used as a numerical algorithm to determine the maximum entropy
state. This can be very useful in practice. Furthermore, these relaxation
equations share interesting analogies with the Debye-H\"uckel equations for
electrolytes \cite{dh2}, the Keller-Segel model of chemotaxis \cite{ks}, and the
Smoluchowski-Poisson system describing self-gravitating Brownian particles
\cite{sopik}.  We have considered a limit of strong mixing (or low energy) where
the maximum entropy state becomes equivalent to a minimum enstrophy state. This
limit is similar to the Debye-H\"uckel \cite{dh1} approximation for
electrolytes, except that the temperature is negative instead of positive so
that the effective interaction between like-sign point vortices is attractive
instead of being repulsive. This results in a self-organization of the system at
the largest scales (condensation), instead of a shielding of the interaction. We
have stressed the limitations of the sinh-Poisson equation that is not the most
general result of the statistical theory. It can miss important solutions such
as the ``monopolar'' or ``spin-up'' state with zero circulation and nonzero
angular momentum.

Point vortices constitute a nice example of systems with long-range interactions that shares numerous analogies with stellar systems, plasmas, and the HMF model \cite{houches}. However, real hydrodynamic flows are not made of point vortices\footnote{An exception concerns quantum vortices in superfluids \cite{onsager}.} contrary to galaxies and plasmas that are basically made of point mass stars or point charges. Therefore, for physical applications \cite{bvrevue}, the point vortex model remains a crude model and one should rather consider the evolution of continuous vorticity flows governed by the 2D Euler equation.\footnote{As explained in the Introduction, the 2D Euler equation also describes the collisionless evolution of the point vortex gas in the $N\rightarrow +\infty$ limit with $\gamma\sim 1/N$.} The 2D Euler equation can reach a statistical equilibrium state as a result of a violent relaxation. This is described by the MRS theory. However, the equilibrium state depends on an infinity of constraints (the Casimirs) in addition to energy. To simplify the problem, Chavanis and Sommeria \cite{jfm1} have considered a strong mixing (or low energy) limit  which makes a hierarchy between these constraints. This amounts to expanding the equations of the MRS theory in powers of $\beta\sigma\psi\ll 1$ (where $\sigma$ denotes the vorticity levels). As the lowest order, only the circulation and the microscopic enstrophy matter, in addition to the energy. When we go to higher orders in the expansion, more and more vorticity moments must be taken into account. For example, at the third order, the Kurtosis plays a crucial role. By contrast, the Kurtosis does not appear in the analogous expansion performed in this paper for point vortices.\footnote{In this analogy, the microscopic moments $\Gamma_n^m=\sum_a N_a\gamma_a^n$ of the vorticity of point vortices are the counterparts of the moments $\Gamma_n^{f.g.}=\int \overline{\omega^n}\, d{\bf r}$  of the fine-grained vorticity.} Therefore, the limit of strong mixing can be used to study the differences between the equilibrium state of point vortices \cite{jm} and the equilibrium state of continuous vorticity field \cite{miller,rs}. This is also a good approach to describe phase transitions in 2D flows with a reduced number of constraints.

\appendix

\section{Density of states and partition function}
\label{sec_heur}

In this Appendix, we justify the variational problems (\ref{m1}) and (\ref{c1}) from the theory of large deviations (without going into technical details) and show their connection to the density of states in the microcanonical ensemble and to the partition function in the canonical ensemble. For the sake of simplicity, we consider a single species gas of point vortices and ignore the conservation of angular momentum and linear impulse (the generalization is straightforward).

\subsection{The microcanonical ensemble}
\label{sec_mcd}

For an isolated Hamiltonian system of point vortices at statistical  equilibrium, the  $N$-body distribution is given by the microcanonical distribution
\begin{equation}
\label{xm1} P_{N}({\bf r}_{1},...,{\bf r}_{N})={1\over
g(E)} \delta [E-H({\bf r}_{1},...,{\bf r}_{N})]
\end{equation}
expressing the equiprobability of the accessible microscopic configurations. This distribution gives the probability density of the microstate  $({\bf r}_1,...,{\bf r}_N)$. This distribution is based on the postulate that all the accessible microscopic configurations (that have energy $E$) are equiprobable. Therefore, it assumes a uniform probability distribution for the phase space variables ${\bf r}_1$, ${\bf r}_2$,... over the energy shell at $H=E$.   The normalization factor is the density of states with energy $E$. It is given by
\begin{equation}
\label{xm2} g(E)=\int \delta [E-H({\bf r}_{1},...,{\bf r}_{N})] \, \prod_{i}d{\bf r}_{i}.
\end{equation}
The number of microstates with energy between $E$ and $E+dE$ is $g(E)dE$.
The microcanonical entropy of the system is defined by
$S(E)= \ln g(E)$ and the microcanonical temperature by
$1/T(E)=\partial S/\partial E$. The entropy is defined up to an additive constant.

We introduce the smooth (coarse-grained) density of point vortices $n({\bf r})$. A microstate is determined by the specification of the
exact positions $\lbrace {\bf r}_i\rbrace$ of the $N$ point vortices. A
macrostate is determined by  the specification of the smooth density
$n({\bf r})$ of point vortices in each cell $[{\bf r},{\bf r}+d{\bf r}]$ irrespectively of their  precise position in the
cell.  Let us call $W[\omega]$ the unconditional number of  microstates $\lbrace{\bf r}_i\rbrace$
corresponding to the macrostate $\omega({\bf r})=\gamma n({\bf r})$. The entropy of the macrostate $\omega$ is defined  by the
Boltzmann formula
\begin{eqnarray}
S[\omega]=\ln W[\omega].
\label{xm4}
\end{eqnarray}
The unconditional probability density of the distribution $\omega$ is therefore $P_0[\omega]\propto W[\omega]\propto e^{S[\omega]}$. The number of complexions  $W[\omega]$ can be obtained by a standard combinatorial analysis. For $N\gg 1$, we find that the Boltzmann entropy is given by
\begin{eqnarray}
\label{xm5}
S[\omega]=-\int \frac{\omega}{\gamma}\ln \left (\frac{\omega}{N\gamma}\right )\, d{\bf r}.
\end{eqnarray}

Instead of integrating over the microstates  $\lbrace {\bf r}_i\rbrace$  in Eq. (\ref{xm2}), we can integrate over the macrostates $\lbrace\omega({\bf r})\rbrace$. For $N\gg 1$, using a mean field approximation, we find that the density of states (\ref{xm2}) can be  written as
\begin{eqnarray}
g(E)\simeq \int  W[\omega] \delta(E[\omega]-E)\delta(\Gamma[\omega]-\Gamma)\, {\cal D}\omega,
\label{xm6}
\end{eqnarray}
where $E[\omega]=\frac{1}{2}\int\omega\psi\, d{\bf r}$ is the mean field energy, $\Gamma[\omega]=\int\omega\, d{\bf r}$ is the circulation, and $W[\omega]$ gives the number of microstates corresponding to the macrostate $\omega$. Since  $W[\omega]= e^{S[\omega]}$, the foregoing expression may be rewritten as
\begin{eqnarray}
g(E)\simeq \int  e^{S[\omega]} \delta(E[\omega]-E)\delta(\Gamma[\omega]-\Gamma)\, {\cal D}\omega.
\label{xm7}
\end{eqnarray}
The microcanonical density probability of the distribution $\omega$ is therefore
\begin{eqnarray}
P[\omega]=\frac{1}{g(E)}e^{S[\omega]}\delta(E[\omega]-E)\delta(\Gamma[\omega]-\Gamma).
\label{xm8}
\end{eqnarray}
This distribution can be obtained directly by stating that $P[\omega]\propto W[\omega]\delta(E[\omega]-E)\delta(\Gamma[\omega]-\Gamma)$ since all the accessible microstates are equiprobable.

\subsection{Canonical ensemble}

For a system of Brownian point vortices in contact with a thermal bath at statistical  equilibrium \cite{pvbrownian}, the  $N$-body distribution is given by the canonical distribution
\begin{eqnarray}
P_{N}({\bf r}_{1},...,{\bf r}_{N})=\frac{1}{Z(\beta)}e^{-\beta H({\bf r}_{1},...,{\bf r}_{N})}.
\label{nc1}
\end{eqnarray}
This distribution gives the probability density of the microstate  $({\bf r}_1,...,{\bf r}_N)$.   The normalization factor is the partition function. It is given by
\begin{eqnarray}
Z(\beta)=\int e^{-\beta H({\bf r}_{1},...,{\bf r}_{N})} \, d{\bf r}_1...d{\bf r}_N.
\label{nc2}
\end{eqnarray}
The free energy is defined by $J(\beta)=\ln Z(\beta)$. The average energy $E=\langle H\rangle$ is given by $E=-\partial J/\partial\beta$. The fluctuations of energy are
given by $\langle H^2\rangle-\langle H\rangle^2=T^2 C$ where $C=dE/dT$ is
the specific heat. This relation implies that the specific heat is always
positive in the canonical ensemble.

Instead of integrating over the microstates  $\lbrace {\bf r}_i\rbrace$  in Eq. (\ref{nc2}), we can integrate over the macrostates $\lbrace\omega({\bf r})\rbrace$. Introducing  the unconditional number of microstates $W[\omega]$ corresponding to the macrostate $\omega$, and using a mean field approximation valid for $N\gg 1$, we find that the partition function can be written as
\begin{eqnarray}
Z(\beta) \simeq  \int e^{-\beta E[\omega]} \, W[\omega] \, \delta(\Gamma[\omega]-\Gamma)\, {\cal D}\omega\nonumber\\
\simeq \int  e^{S[\omega]-\beta E[\omega]} \, \delta(\Gamma[\omega]-\Gamma)\, {\cal D}\omega\nonumber\\
\simeq \int  e^{J[\omega]} \, \delta(\Gamma[\omega]-\Gamma)\, {\cal D}\omega,
\label{nc4}
\end{eqnarray}
where $J[\omega]=S[\omega]-\beta S[\omega]$ is the mean field  free energy.  The canonical density probability of the distribution $\omega$ is therefore
\begin{eqnarray}
P[\omega]=\frac{1}{Z(\beta)}e^{J[\omega]}\delta(\Gamma[\omega]-\Gamma).
\label{nc5}
\end{eqnarray}
This distribution can be obtained directly by stating that $P[\omega]\propto W[\omega] e^{-\beta  E[\omega]}\delta(\Gamma[\omega]-\Gamma)$ since the microstates with energy $E$ have a probability $\propto e^{-\beta E}$.

\subsection{Variational principles}

In the thermodynamic limit $N\rightarrow +\infty$ defined in Sec. \ref{sec_mfap}, we have the extensive scalings $S\sim E/T\sim N$ and $J\sim N$, so we can define  $S[\omega]=Ns[\omega]$ and  $J[\omega]=Nj[\omega]$ with $s\sim 1$ and $j\sim 1$. Accordingly, the preceding results may be  rewritten as
\begin{eqnarray}
g(E)\simeq \int e^{N s[\omega]}\,
\delta(E[\omega]-E)\delta(\Gamma[\omega]-\Gamma) \, {\cal D}\omega
\label{v1}
\end{eqnarray}
and
\begin{eqnarray}
Z(\beta)\simeq \int  e^{ N j[\omega]}\, \delta(\Gamma[\omega]-\Gamma) \, {\cal D}\omega,
\label{v2}
\end{eqnarray}
where the $\delta$-functions take into account the constraints in the microcanonical and canonical ensembles.

The microcanonical probability of the macrostate $\omega$ is
\begin{equation}
P[\omega]=\frac{1}{g(E)}e^{N s[\omega]}\,
\delta(E[\omega]-E)\delta(\Gamma[\omega]-\Gamma)
\end{equation}
and the canonical probability of the macrostate $\omega$ is
\begin{eqnarray}
P[\omega]=\frac{1}{Z(\beta)}e^{Nj[\omega]}\, \delta(\Gamma[\omega]-\Gamma).
\end{eqnarray}

For $N\rightarrow +\infty$, we can make the saddle point approximation. In the microcanonical ensemble, we obtain
\begin{eqnarray}
g(E)=e^{S(E)}\simeq e^{N s[\omega_*]},
\label{heur20}
\end{eqnarray}
i.e.
\begin{eqnarray}
\lim_{N\rightarrow +\infty} \frac{1}{N}S(E)=s[\omega_*],
\label{heur21}
\end{eqnarray}
where $\omega_*$ is the global maximum of constrained entropy. We are led therefore to the maximization problem
\begin{eqnarray}
S(E)=\max_{\omega}\lbrace S[\omega]\, |\, E[\omega]=E, \, \Gamma[\omega]=\Gamma \rbrace.
\label{heur22}
\end{eqnarray}
At equilibrium, in the $N\rightarrow +\infty$ limit, we have $S(E)=S[\omega_{*}]=N s[\omega_{*}]$. In the canonical ensemble, we obtain
\begin{eqnarray}
Z(\beta)=e^{J(\beta)}\simeq e^{ N j[\omega_*]},
\label{heur23}
\end{eqnarray}
i.e.
\begin{eqnarray}
\lim_{N\rightarrow +\infty} \frac{1}{N}J(\beta)=j[\omega_*],
\label{heur24}
\end{eqnarray}
where $\omega_*$ is the global maximum of constrained free energy. We are led therefore to the maximization problem
\begin{eqnarray}
J(\beta)=\max_{\omega}\lbrace J[\omega]\, |\, \Gamma[\omega]=\Gamma   \rbrace.
\label{heur25}
\end{eqnarray}
At equilibrium, in the $N\rightarrow +\infty$ limit, we have $J(\beta)=N J[\omega_*]=N j[\omega_{*}]$.

These results can be considered as results of large deviations \cite{ellisld,touchette} and they can be made mathematically rigorous \cite{caglioti1,kiessling,eyink,caglioti2,kl,sawada}.

\section{Uniformly rotating or translating steady states of the 2D Euler equation}
\label{sec_euler}

In the $N\rightarrow +\infty$ limit with $\gamma\sim 1/N$, the average vorticity
$\omega({\bf r},t)=\langle \sum_i \gamma_i \delta({\bf r}-{\bf r}_i(t))\rangle$
of the point vortex gas is governed by the 2D Euler equation
\begin{eqnarray}
\frac{\partial\omega}{\partial t}+{\bf u}\cdot \nabla\omega=0, \qquad \omega=-\Delta\psi.
\label{euler1}
\end{eqnarray}

If the flow is steady in a frame rotating with angular velocity $\Omega$, then $\omega(r,\theta,t)=\omega(r,\theta-\Omega t)$ where $(r,\theta)$ is a polar system of coordinates. This implies that $\partial_t\omega=-\Omega\partial_{\theta}\omega$. Substituting this relation in the 2D Euler equation we obtain
\begin{eqnarray}
\frac{\partial}{\partial\theta}(\psi+\frac{\Omega}{2}r^2)\frac{\partial\omega}{\partial r}-\frac{\partial}{\partial r}(\psi+\frac{\Omega}{2}r^2)\frac{\partial\omega}{\partial \theta}=0.
\label{euler2}
\end{eqnarray}
This is the general equation determining a steady state in a rotating frame. This equation is satisfied by any relation of the form $\omega=f(\psi+\Omega r^2/2)$.

If the flow is steady in a frame translating with linear velocity ${\bf U}=U {\bf e}_x$, then $\omega(x,y,t)=\omega(x-Ut,y)$ where $(x,y)$ is a cartesian system of coordinates. This implies that $\partial_t\omega=-U\partial_{x}\omega$. Substituting this relation in the 2D Euler equation  we obtain
\begin{eqnarray}
\frac{\partial}{\partial y}(\psi-U y)\frac{\partial\omega}{\partial x}-\frac{\partial}{\partial x}(\psi-U y)\frac{\partial\omega}{\partial y}=0.
\label{euler2b}
\end{eqnarray}
This is the general equation determining a steady state in a translating frame. It is satisfied by any relation of the form $\omega=f(\psi-U y)$.

\section{The Debye length}
\label{sec_debye}

We consider a neutral plasma at equilibrium with uniform charge
density ($\rho=\sum_a n_a e_a=0$). We introduce a test charge $+e$ in the system
at ${\bf r}_P={\bf 0}$. This charge produces a ``naked'' electric potential
$\Phi_{0}({\bf r})$. This potential modifies the distribution of the field
charges around the test charge. The resulting change of density $\tilde\rho({\bf
r})$ in turn produces an extra-potential which adds to the original one
$\Phi_{0}({\bf r})$. The effective, or ``dressed'', potential created by the
test charge is therefore the solution of the  equation
\begin{eqnarray}
\Delta\Phi_{eff}=-S_d \lbrack \tilde\rho({\bf r})+e\delta({\bf r})\rbrack.
\label{debye1}
\end{eqnarray}
The slightly perturbed distribution of the field charges is given by the Boltzmann statistics
\begin{eqnarray}
\tilde\rho({\bf r})=\sum_a n_a e_a e^{-\beta e_a\Phi_{eff}({\bf r})},
\label{debye2}
\end{eqnarray}
where $n_a$ is the uniform distribution of species $a$. Substituting Eq. (\ref{debye2}) in Eq. (\ref{debye1}) we obtain the self-consistency equation
\begin{eqnarray}
\Delta\Phi_{eff}=-S_d \sum_a n_a e_a e^{-\beta e_a\Phi_{eff}}-S_d e\delta({\bf
r})
\label{debye3}
\end{eqnarray}
determining the effective (dressed) potential. In the weak coupling approximation $\beta e_a \Phi_{eff}\ll 1$, we can expand the exponential term in Eq. (\ref{debye3}) leading to the equation
\begin{eqnarray}
\Delta\Phi_{eff}-S_d \beta\left (\sum_a n_a e_a^2\right ) \Phi_{eff}=-S_d e\delta({\bf r}).
\label{debye4}
\end{eqnarray}
This equation shows that the potential created by the test charge is shielded on
a typical length $\lambda_D=(k_B T/S_d\sum_a n_a e_a^2)^{1/2}$ called the Debye
length. This length can also be obtained in a less rigorous manner by expanding
Eq. (\ref{conn6}) for $\beta e_a \Phi\ll 1$.

\section{The $\beta\gamma_a\psi\ll 1$ expansion for point vortices}
\label{sec_smpv}

In this Appendix, we expand the equations of the statistical theory of point
vortices in terms of $\beta\gamma_a\psi\ll 1$ to third order. This generalizes
the study of Sec. \ref{sec_dh} that was limited to second order.

It is convenient to work in terms of the variable $\phi=\psi-\langle\psi\rangle$, satisfying $\langle\phi\rangle=0$, instead of $\psi$.  Eq. (\ref{m3}) is thus rewritten as
\begin{eqnarray}
\omega_a=A_a\gamma_a e^{-\beta\gamma_a\phi},\qquad A_a=\frac{N_a}{\int e^{-\beta\gamma_a\phi}\, d{\bf r}}.
\label{smpv1}
\end{eqnarray}
To third order, we obtain
\begin{eqnarray}
e^{-\beta\gamma_a\phi}=1-\beta\gamma_a\phi+\frac{1}{2}\beta^2\gamma_a^2\phi^2
-\frac{1}{6}\beta^3\gamma_a^3\phi^3 
\label{smpv2}
\end{eqnarray}
and
\begin{eqnarray}
A_a=N_a \left (1-\frac{1}{2}\beta^2\gamma_a^2\langle\phi^2\rangle+\frac{1}{6}\beta^3\gamma_a^3\langle\phi^3\rangle\right ),
\label{smpv3}
\end{eqnarray}
where we have assumed that the domain area is unity ($V=1$). Therefore
\begin{eqnarray}
\omega_a=N_a\gamma_a\biggl\lbrack
1-\beta\gamma_a\phi+\frac{1}{2}\beta^2\gamma_a^2(\phi^2-\langle\phi^2\rangle)\nonumber\\
-\frac{1}{6}\beta^3\gamma_a^3(\phi^3-\langle\phi^3\rangle)
+\frac{1}{2}\beta^3\gamma_a^3\langle\phi^2\rangle \phi\biggr \rbrack.
\label{smpv4}
\end{eqnarray}
Summing over the species, we obtain the total vorticity
\begin{eqnarray}
\label{smpv5}
\omega=\Gamma+C_1\beta\phi+C_2\beta^2(\phi^2-\langle\phi^2\rangle)
+C_3\beta^3(\phi^3-\langle\phi^3\rangle)\nonumber\\
\end{eqnarray}
with
\begin{eqnarray}
\label{smpv6}
C_1=-\Gamma_2^m+\frac{1}{2}\Gamma_4^m\beta^2\langle\phi^2\rangle,
\end{eqnarray}
\begin{eqnarray}
\label{smpv7}
C_2=\frac{1}{2}\Gamma_3^m,\qquad C_3=-\frac{1}{6}\Gamma_4^m,
\end{eqnarray}
where $\Gamma_n^m=\sum_a N_a\gamma_a^n$ denotes the microscopic moment of order
$n$ of the vorticity.
Using Eq. (\ref{smpv4}), the entropy (\ref{b2}) is given  to third order by
\begin{eqnarray}
\label{smpv8}
S=S_0-\frac{1}{2}\Gamma_2^m\beta^2\langle\phi^2\rangle+\frac{1}{3}\Gamma_3^m\beta^3\langle\phi^3\rangle,
\end{eqnarray}
where $S_0=\sum_a N_a\ln N_a$.

We can use this expansion to show the relation between maximum entropy states and minimum enstrophy states in the strong mixing limit. At second order, we get
\begin{eqnarray}
\label{smpvnew1}
\omega=\Gamma-\Gamma_2^{m}\beta\phi+\frac{1}{2}\Gamma_3^{m}\beta^2(\phi^2
-\langle\phi^2\rangle)
\end{eqnarray}
and
\begin{eqnarray}
\label{smpvnew2}
S=S_0-\frac{1}{2}\Gamma_2^{m}\beta^2\langle\phi^2\rangle.
\end{eqnarray}
Introducing the macroscopic enstrophy  $\Gamma_2=\int \omega^2\, d{\bf r}$, and using Eq. (\ref{smpvnew1}), we obtain at second order
\begin{eqnarray}
\label{smpvnew3}
\Gamma_2=\Gamma^2+(\Gamma_2^{m})^2\beta^2\langle\phi^2\rangle.
\end{eqnarray}
Comparing Eq. (\ref{smpvnew2}) with Eq. (\ref{smpvnew3}) we get
\begin{eqnarray}
\label{smpvnew4}
S=S_0-\frac{1}{2}\frac{\Gamma_2-\Gamma^2}{\Gamma_2^{m}}.
\end{eqnarray}
This relation shows that, at second order in the strong mixing limit, a maximum entropy state is equivalent to a minimum enstrophy state.

In the case of a symmetric distribution of point vortices with positive and negative circulations, we have $\Gamma_{2n+1}^m=0$, so Eqs. (\ref{smpv5})-(\ref{smpv7}) reduce to
\begin{eqnarray}
\label{smpv9}
\omega=\left (-\Gamma_2^m+\frac{1}{2}(\Gamma_2^m)^2{\rm Ku}\,\beta^2\langle\phi^2\rangle\right )\beta\phi\nonumber\\
-\frac{1}{6}(\Gamma_2^m)^2{\rm Ku}\, \beta^3(\phi^3-\langle\phi^3\rangle),
\end{eqnarray}
where ${\rm Ku}=\Gamma_4^{m}/(\Gamma_2^{m})^2$ is the Kurtosis of the microscopic  vorticity distribution. On the other hand, the entropy (\ref{smpv8}) reduces to
\begin{eqnarray}
\label{smpv8b}
S=S_0-\frac{1}{2}\beta^2\Gamma_2^m\langle\phi^2\rangle.
\end{eqnarray}

We now consider the low energy limit $E\ll 1$ (still assuming
$\Gamma_{2n+1}^m=0$) and use standard perturbation theory \cite{taylor,ashbee}.
We write $\phi=E^{1/2}(\phi_0+E\phi_1+...)$ and $\beta=\beta_0+E\beta_1+...$ and
substitute these expansions in Eq. (\ref{smpv9}). At order $E^{1/2}$ we get
\begin{eqnarray}
\label{smpv10}
-\Delta\phi_0+\Gamma_2^m\beta_0\phi_0=0,
\end{eqnarray}
and at order $E^{3/2}$ we obtain
\begin{eqnarray}
\label{smpv11}
-\Delta\phi_1+\Gamma_2^m\beta_0\phi_1=-\Gamma_2^m\beta_1\phi_0
+\frac{1}{2}(\Gamma_2^m)^2 {\rm Ku}\, \beta_0^3\langle\phi_0^2\rangle\phi_0\nonumber\\
-\frac{1}{6}(\Gamma_2^m)^2 {\rm Ku}\, \beta_0^3(\phi_0^3-\langle\phi_0^3\rangle).\qquad 
\end{eqnarray}
The linear problem (\ref{smpv10}) has been studied in \cite{jfm1} in terms of
$\psi$. The inverse temperature $\beta_0$ is ``quantized'' and can take only
discrete values $\beta_n<0$ labeled by $n=1,2,...$. When
$\langle\psi_0\rangle=0$ (type I solutions), $\Gamma_2^m\beta_0$ is an
eigenvalue of the Laplacian $\Delta$ with zero mean and with the boundary
condition $\psi_0=0$ on $(\partial{\cal D})$. When $\langle\psi_0\rangle\neq 0$
(type II solutions), $\Gamma_2^m\beta_0$ is a root of the function $F(\beta)$
defined by Eq. (3.8) of \cite{jfm1} (it is constructed with the eigenvalues of
the Laplacian with non-zero mean and with the boundary condition $\psi_0=0$ on
$(\partial{\cal D})$). The respective values of  $\beta_0$ for type I and type
II solutions depend on the shape of the domain. It is the competition between
these two types of solutions that yields the geometry-induced phase transitions
found in \cite{jfm1}. Coming back to the variable $\phi$, we call $\Psi_n$ the
ortho-normalized basis of eigenfunctions associated with the eigenvalues
$\Gamma_2^m\beta_n$ regrouping the two types of solutions described above. They
are defined by $-\Delta\Psi_n+\Gamma_2^m\beta_n\Psi_n=0$ with $\Psi_n={\rm
cst.}$ on the boundary (not necessarily zero), $\langle\Psi_n\rangle=0$, and
$\langle \Psi_n|\Psi_{n'}\rangle=\delta_{nn'}$ where $\langle f|g\rangle=\int
fg\, d{\bf r}$.  The solution of Eq. (\ref{smpv10}) satisfying the
energy constraint $E=\frac{1}{2}\int\omega\psi\, d{\bf
r}=\frac{1}{2}\int\omega\phi\, d{\bf r}=-\frac{1}{2}\int\phi\Delta\phi\, d{\bf
r}$ (since $\Gamma=0$) can be written as
\begin{eqnarray}
\label{smpv12}
\phi_0^{(n)}=\left (\frac{-2}{\beta_n\Gamma_2^m}\right )^{1/2}\Psi_n,\qquad \beta_0^{(n)}=\beta_n.
\end{eqnarray}
This defines the starting point (at $E=0$) of the branch $\beta^{(n)}(E)$ of
order $n$. In order 
to get the first order correction to the temperature, we multiply Eq.
(\ref{smpv11}) by $\phi_0^{(n)}$ and integrate over the domain. Using the
identity
\begin{eqnarray}
\label{smpv13}
\langle \phi_0|-\Delta\phi_1+\Gamma_2^m\beta_0\phi_1\rangle= \langle -\Delta\phi_0+\Gamma_2^m\beta_0\phi_0|\phi_1\rangle=0\nonumber\\
\end{eqnarray}
which results from a simple integration by parts (we have omitted the superscript $(n)$ on the variables to simplify the expression), we obtain
\begin{eqnarray}
\label{smpv14}
\beta_1^{(n)}={\rm Ku}\, \beta_n^2 \left (\frac{1}{3}\langle\Psi_n^4\rangle-1\right ).
\end{eqnarray}
This result, which can be viewed as a solvability condition, generalizes 
the perturbative result obtained by Taylor {\it et al.} \cite{taylor} (see
also \cite{ashbee}) for a two species system of point vortices $(N/2,\gamma)$
and $(N/2,-\gamma)$ for which ${\rm
Ku}=\Gamma_4^m/(\Gamma_2^m)^2=N\gamma^4/(N\gamma^2)^2=1/N$. On the other hand,
the first order correction to the stream function can be expanded on the basis
of eigenfunctions as
\begin{eqnarray}
\label{smpv15}
\phi_1^{(n)}=\sum_k c_k^{(n)} \Psi_k,\qquad c_k^{(n)}=\langle \Psi_k|\phi_1^{(n)}\rangle.
\end{eqnarray}
Multiplying Eq. (\ref{smpv11}) by $\Psi_k$ with  $k\neq n$, we find
\begin{eqnarray}
\label{smpv16}
c_k^{(n)}=\frac{1}{3}\beta_n^2 \left (\frac{-2}{\beta_n\Gamma_2^m}\right )^{1/2}{\rm Ku}\frac{\langle\Psi_k|\Psi_n^3\rangle}{\beta_n-\beta_k}.
\end{eqnarray}
The coefficient $c_n^{(n)}$ can be taken equal to zero.

The specific heat is defined by $C=dE/dT=-\beta^2 dE/d\beta$. For $E\rightarrow
0$ we obtain $C=-\beta_n^2/\beta_1^{(n)}$ where $\beta_1^{(n)}$ is given by Eq.
(\ref{smpv14}). The specific heat is positive when $\langle\Psi_n^4\rangle<3$,
negative when $\langle\Psi_n^4\rangle>3$, and infinite when
$\langle\Psi_n^4\rangle=3$. Depending on the respective values of
$\beta_1^{(n)}$, the different branches $n=1,2...$ may cross each other at some
energy $E_c>0$ leading to energy-induced phase transitions (their crossing may
also be due to nonlinear effects not captured by the perturbative expansion)
\cite{taylor,ashbee}.

At order $E^{3/2}$, using Eq. (\ref{smpv8b}) and
$E=-\frac{1}{2}\Gamma_2^m\beta\langle\phi^2\rangle$ we obtain
\begin{eqnarray}
\label{smpv17}
S^{(n)}=S_0+\beta_n E.
\end{eqnarray}
At $E=0$, all the solutions $n=1,2...$ have the same entropy $S_0$.
For $0<E\ll 1$, the entropies $S^{(n)}$ of the different solutions are just
proportional to their inverse temperature $\beta_n$.  Therefore, the maximum
entropy state is the mode $n=1$, i.e. the one with the highest $\beta_n$
\cite{jfm1}. If two
modes have the same inverse temperature ($\beta_n=\beta_m$ with $n\neq m$),
there is a degeneracy that has to be raised by expanding the entropy at order
$E^{2}$.

\section{The $\beta\sigma\psi\ll 1$ expansion for continuous vorticity fields}
\label{sec_smcv}

In this Appendix, we expand the equations of the MRS statistical theory for continuous vorticity fields in terms of $\beta\sigma\psi\ll 1$ to third order. This complements the study of Chavanis and Sommeria \cite{jfm1} where this expansion was developed in detail to second order and extended (without giving detail) to third order. This Appendix also clarifies the analogies  and the differences between the statistical mechanics of point vortices and continuous vorticity fields. We refer to \cite{jfm1} for a detailed presentation of the MRS theory and for the notations. Below, we just recall the basic formulae that are needed for our study.  

As in Appendix \ref{sec_smpv}, it is convenient to work in terms of the variable $\phi=\psi-\langle\psi\rangle$, satisfying $\langle\phi\rangle=0$, instead of $\psi$. According to the MRS theory, the probability density of finding the vorticity level $\sigma$ in ${\bf r}$ is given by
\begin{eqnarray}
\rho({\bf r},\sigma)=\frac{1}{Z({\bf r})}g(\sigma)e^{-\beta\sigma\phi}
\label{smcv1}
\end{eqnarray}
with the normalization condition $\int\rho({\bf r},\sigma)\, d\sigma=1$ leading
to
\begin{eqnarray}
Z({\bf r})=\int g(\sigma)e^{-\beta\sigma\phi}\, d\sigma.
\label{smcv2}
\end{eqnarray}
The function $g(\sigma)$ can be viewed as a Lagrange multiplier determined by
the total area $\gamma(\sigma)=\int \rho({\bf r},\sigma)\, d{\bf r}$ of each
vorticity level $\sigma$ which is a conserved quantity.  The coarse-grained
vorticity field is given by $\overline{\omega}({\bf r})=\int\rho({\bf
r},\sigma)\sigma\, d\sigma$. The entropy is
\begin{eqnarray}
\label{smcv2b}
S=-\int \rho({\bf r},\sigma)\ln\rho({\bf r},\sigma)\, d{\bf r}d\sigma.
\end{eqnarray}

To third order, we obtain
\begin{eqnarray}
e^{-\beta\sigma\phi}=1-\beta\sigma\phi+\frac{1}{2}\beta^2\sigma^2\phi^2
-\frac{1}{6}\beta^3\sigma^3\phi^3
\label{smcv3}
\end{eqnarray}
and
\begin{eqnarray}
Z=1-\beta A_1\phi+\frac{1}{2}\beta^2 A_2\phi^2
-\frac{1}{6}\beta^3 A_3\phi^3,
\label{smcv4}
\end{eqnarray}
where we have defined $A_n=\int g(\sigma)\sigma^n\, d\sigma$ and taken $A_0=1$ without loss of generality. Therefore,
\begin{eqnarray}
\rho({\bf r},\sigma)=g(\sigma)\biggl\lbrack 1+\beta\phi(A_1-\sigma)\nonumber\\
+\beta^2\phi^2\left (\frac{1}{2}\sigma^2-A_1\sigma-\frac{1}{2}A_2+A_1^2\right )
+\beta^3\phi^3\biggl (-\frac{1}{6}\sigma^3\nonumber\\
+\frac{1}{2}A_1\sigma^2
+\frac{1}{2}A_2\sigma+\frac{1}{6}A_3-A_1^2\sigma-A_1A_2+A_1^3\biggr )\biggr\rbrack.\nonumber\\
\label{smcv5}
\end{eqnarray}
Integrating Eq. (\ref{smcv5}) over the whole domain, we obtain $\gamma(\sigma)$
as a function of $g(\sigma)$. Reversing this relation, we obtain
\begin{eqnarray}
g(\sigma)=\gamma(\sigma)\biggl\lbrack 1
-\beta^2\langle \phi^2\rangle\left (\frac{1}{2}\sigma^2-A_1\sigma-\frac{1}{2}A_2+A_1^2\right )\nonumber\\
-\beta^3\langle \phi^3\rangle \biggl (-\frac{1}{6}\sigma^3
+\frac{1}{2}A_1\sigma^2
+\frac{1}{2}A_2\sigma\nonumber\\
+\frac{1}{6}A_3-A_1^2\sigma-A_1A_2+A_1^3\biggr )\biggr\rbrack,\nonumber\\
\label{smcv6}
\end{eqnarray}
where we have assumed that the domain area is unity ($V=1$). We note that $g(\sigma)$ still appears implicitly in the coefficients $A_n$. Multiplying Eq. (\ref{smcv6}) by $\sigma^n$ and integrating over $\sigma$, we get
\begin{eqnarray}
A_n=\Gamma_n^{f.g.}-\beta^2\langle \phi^2\rangle\biggl (\frac{1}{2}\Gamma_{n+2}^{f.g.}-A_1\Gamma_{n+1}^{f.g.}
-\frac{1}{2}A_2\Gamma_n^{f.g.}\nonumber\\
+A_1^2\Gamma_n^{f.g.}\biggr )
-\beta^3\langle \phi^3\rangle \biggl (-\frac{1}{6}\Gamma_{n+3}^{f.g.}
+\frac{1}{2}A_1\Gamma_{n+2}^{f.g.}
+\frac{1}{2}A_2\Gamma_{n+1}^{f.g.}\nonumber\\
+\frac{1}{6}A_3\Gamma_n^{f.g.}
-A_1^2\Gamma_{n+1}^{f.g.}-A_1A_2\Gamma_n^{f.g.}+A_1^3\Gamma_n^{f.g.}\biggr )\biggr\rbrack,\nonumber\\
\label{smcv7}
\end{eqnarray}
where $\Gamma_n^{f.g.}=\int \overline{\omega^n}\, d{\bf r}=\int \gamma(\sigma)\sigma^n\, d\sigma$ is the moment of order $n$ of the fine-grained vorticity field (conserved quantity). Identifying terms of equal order in Eq. (\ref{smcv7}) we obtain at second order
\begin{eqnarray}
A_n=\Gamma_n^{f.g.}-\beta^2\langle \phi^2\rangle\biggl (\frac{1}{2}\Gamma_{n+2}^{f.g.}-\Gamma\Gamma_{n+1}^{f.g.}\nonumber\\
-\frac{1}{2}\Gamma_2^{f.g.}\Gamma_n^{f.g.}
+\Gamma^2\Gamma_n^{f.g.}\biggr ).
\label{smcv8}
\end{eqnarray}
Combining Eqs. (\ref{smcv5}), (\ref{smcv6}), and (\ref{smcv8}) we get
\begin{eqnarray}
\label{smcv8b}
\rho({\bf r},\sigma)=\gamma(\sigma)\biggl\lbrack 1+B_1\beta\phi+B_2\beta^2(\phi^2-\langle\phi^2\rangle)\nonumber\\
+B_3\beta^3(\phi^3-\langle\phi^3\rangle)\biggr\rbrack,
\end{eqnarray}
with
\begin{eqnarray}
\label{smcv9}
B_1=\Gamma-\sigma-\beta^2\langle\phi^2\rangle \biggl ( \frac{1}{2}\Gamma_3^{f.g.}-\frac{3}{2}\Gamma\Gamma_2^{f.g.}
+2\Gamma^3\nonumber\\
+\frac{3}{2}\Gamma\sigma^2
-2\Gamma^2\sigma-\frac{1}{2}\Gamma\Gamma_2^{f.g.}
-\frac{1}{2}\sigma^3+\frac{1}{2}\Gamma_2^{f.g.}\sigma\biggr ),
\end{eqnarray}
\begin{eqnarray}
\label{smcv10}
B_2=\frac{1}{2}\sigma^2-\Gamma\sigma-\frac{1}{2}\Gamma_2^{f.g.}+\Gamma^2,
\end{eqnarray}
\begin{eqnarray}
\label{smcv11}
B_3=-\frac{1}{6}\sigma^3+\frac{1}{2}\Gamma\sigma^2+\frac{1}{2}\Gamma_2^{f.g.}\sigma
+\frac{1}{6}\Gamma_3^{f.g.}\nonumber\\
-\Gamma^2\sigma-\Gamma\Gamma_2^{f.g.}+\Gamma^3.
\end{eqnarray}
The coarse-grained vorticity is
\begin{eqnarray}
\label{smcv12}
\overline{\omega}=\Gamma+C_1\beta\phi+C_2\beta^2(\phi^2-\langle\phi^2\rangle)
+C_3\beta^3(\phi^3-\langle\phi^3\rangle)\nonumber\\
\end{eqnarray}
with
\begin{eqnarray}
\label{smcv13}
C_1=\Gamma^2-\Gamma_2^{f.g.}-\beta^2\langle\phi^2\rangle \biggl \lbrack 2\Gamma\Gamma_3^{f.g.}-4\Gamma^2\Gamma_2^{f.g.}\nonumber\\
+2\Gamma^4-\frac{1}{2}\Gamma_4^{f.g.}
+\frac{1}{2}(\Gamma_2^{f.g.})^2\biggr \rbrack,
\end{eqnarray}
\begin{eqnarray}
\label{smcv14}
C_2=\frac{1}{2}(\Gamma_3^{f.g.}-3\Gamma\Gamma_2^{f.g.}+2\Gamma^3),
\end{eqnarray}
\begin{eqnarray}
\label{smcv15}
C_3=-\frac{1}{6}(\Gamma_4^{f.g.}-3(\Gamma_2^{f.g.})^2-4\Gamma_3^{f.g.}\Gamma\nonumber\\
+12\Gamma^2\Gamma_2^{f.g.}
-6\Gamma^4).
\end{eqnarray}
These coefficients can be related to the cumulants of
the generating funtion $\ln Z(\phi)$. Using Eq. (\ref{smcv8b}),  the entropy
(\ref{smcv2b}) is given to third order by
\begin{eqnarray}
\label{smcv16}
S=S_0-\frac{1}{2}(\Gamma_2^{f.g.}-\Gamma^2)\beta^2\langle\phi^2\rangle\nonumber\\
+\frac{1}{6}(4\Gamma^3-6\Gamma\Gamma_2^{f.g.}+2\Gamma_3^{f.g.})\beta^3\langle\phi^3\rangle,
\end{eqnarray}
where $S_0=-\int \gamma(\sigma)\ln\gamma(\sigma)\, d\sigma$.

We can use this expansion to show the relation between maximum entropy states and minimum enstrophy states in the strong mixing limit. At second order, we get
\begin{eqnarray}
\label{smcv17}
\overline{\omega}=\Gamma+(\Gamma^2-\Gamma_2^{f.g.})\beta\phi\nonumber\\
+\frac{1}{2}(\Gamma_3^{f.g.}-3\Gamma\Gamma_2^{f.g.}+2\Gamma^3)\beta^2(\phi^2
-\langle\phi^2\rangle)
\end{eqnarray}
and
\begin{eqnarray}
\label{smcv18}
S=S_0-\frac{1}{2}(\Gamma_2^{f.g.}-\Gamma^2)\beta^2\langle\phi^2\rangle.
\end{eqnarray}
Introducing the enstrophy of the coarse-grained field $\Gamma_2=\int\overline{\omega}^2\, d{\bf r}$, and using Eq. (\ref{smcv17}), we obtain at second order
\begin{eqnarray}
\label{smcv19}
\Gamma_2=\Gamma^2+(\Gamma^2-\Gamma_2^{f.g.})^2\beta^2\langle\phi^2\rangle.
\end{eqnarray}
Comparing Eq. (\ref{smcv18}) with Eq. (\ref{smcv19}) we obtain
\begin{eqnarray}
\label{smcv20}
S=S_0-\frac{1}{2}\frac{\Gamma_2-\Gamma^2}{\Gamma_2^{f.g.}-\Gamma^2}.
\end{eqnarray}
Since $\Gamma_2^{f.g.}-\Gamma^2>0$, this relation shows that, at second order in the strong mixing limit, a maximum entropy state is equivalent to a minimum enstrophy state.

In the case of a symmetric distribution of vorticity levels, such that $\Gamma_{2n+1}^{f.g.}=0$, Eqs. (\ref{smcv12})-(\ref{smcv15}) reduce to
\begin{eqnarray}
\label{smcv21}
\overline{\omega}=\left\lbrack -\Gamma_2^{f.g.}+\frac{1}{2}(\Gamma_2^{f.g.})^2({\rm Ku}-1)\beta^2\langle\phi^2\rangle\right\rbrack\beta\phi\nonumber\\
-\frac{1}{6}(\Gamma_2^{f.g.})^2({\rm Ku}-3)\beta^3(\phi^3-\langle\phi^3\rangle),
\end{eqnarray}
where ${\rm Ku}=\Gamma_4^{f.g.}/(\Gamma_2^{f.g.})^2$ is the Kurtosis of the microscopic  vorticity distribution. On the other hand, the entropy (\ref{smcv16}) reduces to
\begin{eqnarray}
\label{smcv22}
S=S_0-\frac{1}{2}\Gamma_2^{f.g.}\beta^2\langle\phi^2\rangle.
\end{eqnarray}

We now consider the low energy limit $E\ll 1$ (still assuming $\Gamma_{2n+1}^{f.g.}=0$) and use standard perturbation theory. We write $\phi=E^{1/2}(\phi_0+E\phi_1+...)$ and $\beta=\beta_0+E\beta_1+...$ and substitute these expansions in Eq. (\ref{smcv21}). At order $E^{1/2}$ we get
\begin{eqnarray}
\label{smcv23}
-\Delta\phi_0+\Gamma_2^{f.g.}\beta_0\phi_0=0,
\end{eqnarray}
and at order $E^{3/2}$ we obtain
\begin{eqnarray}
\label{smcv24}
-\Delta\phi_1+\Gamma_2^{f.g.}\beta_0\phi_1=-\Gamma_2^{f.g.}\beta_1\phi_0\nonumber\\
+\frac{1}{2}(\Gamma_2^{f.g.})^2({\rm Ku}-1)\beta_0^3\langle\phi_0^2\rangle\phi_0\nonumber\\
-\frac{1}{6}(\Gamma_2^{f.g.})^2({\rm Ku}-3)\beta_0^3(\phi_0^3-\langle\phi_0^3\rangle).
\end{eqnarray}
Proceeding as in Appendix \ref{sec_smpv}, the solution of Eq. (\ref{smcv23}) satisfying the energy constraint  is
\begin{eqnarray}
\label{smcv25}
\phi_0^{(n)}=\left (\frac{-2}{\beta_n\Gamma_2^{f.g.}}\right )^{1/2}\Psi_n,\qquad \beta_0^{(n)}=\beta_n.
\end{eqnarray}
The solvability condition gives the first correction to the inverse temperature
\begin{eqnarray}
\label{smcv26}
\beta_1^{(n)}=\beta_n^2\left \lbrack \frac{1}{3}({\rm Ku}-3)\langle\Psi_n^4\rangle-({\rm Ku}-1)\right \rbrack.
\end{eqnarray}
On the other hand, the first correction to the stream function is given by
\begin{eqnarray}
\label{smcv27}
\phi_1^{(n)}=\sum_k c_k^{(n)} \Psi_k,\qquad c_k^{(n)}=\langle \Psi_k|\phi_1^{(n)}\rangle,
\end{eqnarray}
with
\begin{eqnarray}
\label{smcv28}
c_k^{(n)}=\frac{1}{3}\beta_n^2 \left (\frac{-2}{\beta_n\Gamma_2^{f.g.}}\right )^{1/2}({\rm Ku}-3)\frac{\langle\Psi_k|\Psi_n^3\rangle}{\beta_n-\beta_k}
\end{eqnarray}
for $k\neq n$ and $c_n^{(n)}=0$. 

The specific heat is defined by $C=dE/dT=-\beta^2 dE/d\beta$. For $E\rightarrow
0$ we obtain $C=-\beta_n^2/\beta_1^{(n)}$ where $\beta_1^{(n)}$ is given by Eq.
(\ref{smcv26}). There is a critical value of the Kurtosis given by
\begin{eqnarray}
\label{smcv31}
({\rm Ku})_{c}^{(n)}=\frac{\langle \Psi_n^4\rangle-1}{\frac{1}{3}\langle \Psi_n^4\rangle-1}.
\end{eqnarray}
If $\langle \Psi_n^4\rangle<1$, the specific heat is negative for ${\rm Ku}<({\rm Ku})_c$ and positive for ${\rm Ku}>({\rm Ku})_c$.  If $1<\langle \Psi_n^4\rangle<3$, the specific heat is always positive. If $\langle \Psi_n^4\rangle>3$, the specific heat is positive for ${\rm Ku}<({\rm Ku})_c$ and negative for ${\rm Ku}>({\rm Ku})_c$.

We note that the results obtained for point vortices (see Appendix
\ref{sec_smpv}) are recovered from the present results obtained for continuous
vorticity fields if we keep only the moment of highest order at each level of
the expansion. This implies, in particular, that the terms ${\rm Ku}-3$ and
${\rm Ku}-1$ for continuous vorticity fields are replaced by ${\rm Ku}$ for
point vortices.\footnote{The equations of the statistical theory of point
vortices can be recovered from the MRS theory in a dilute limit. It corresponds,
for example, to an initial condition involving compact vortices, thus having
high Kurtosis \cite{jfm1}.} On the other hand, the discussion of the entropy and
of the possible degeneracy of the solutions for continuous vorticity fields is
the same as in the case of point vortices (see the last paragraph of Appendix
\ref{sec_smpv} where it suffices to replace $\Gamma_n^{m}$ by
$\Gamma_n^{f.g.}$).

The relaxation equations for the probability density $\rho({\bf r},\sigma,t)$ associated with the MRS theory are \cite{rsmepp,csr}:
\begin{equation}
\frac{\partial\rho}{\partial t}+{\bf u}\cdot \nabla\rho=\nabla\cdot \left\lbrace D({\bf r},t) \left\lbrack \nabla\rho+\beta(t)(\sigma-\overline{\omega})\rho\nabla\psi\right\rbrack\right\rbrace,
\label{smcv31b}
\end{equation}
\begin{equation}
\beta(t)=-\frac{\int D\nabla\overline{\omega}\cdot\nabla\psi\, d{\bf r}}{\int D\omega_2 (\nabla\psi)^2\, d{\bf r}},
\label{smcv32}
\end{equation}
where $\omega_2=\overline{\omega^2}-\overline{\omega}^2=\int \rho
(\sigma-\overline{\omega})^2\, d\sigma$ is the local centered variance of the
vorticity. They can be simplified in the limit of strong mixing (or low energy).
At leading order, it suffices to make the approximation $\rho({\bf
r},\sigma,t)\simeq \gamma(\sigma)$ and $\overline{\omega}({\bf r},t)\simeq
\Gamma$ in the drift term. 
This yields 
\begin{equation}
\frac{\partial\rho}{\partial t}+{\bf u}\cdot \nabla\rho=\nabla\cdot \left\lbrace D({\bf r},t) \left\lbrack \nabla\rho+\beta(t)(\sigma-\Gamma)\gamma(\sigma)\nabla\psi\right\rbrack\right\rbrace,
\label{smcv33}
\end{equation}
\begin{equation}
\beta(t)=-\frac{\int D\nabla\overline{\omega}\cdot\nabla\psi\, d{\bf r}}{(\Gamma_2^{f.g.}-\Gamma^2)\int D (\nabla\psi)^2\, d{\bf r}}.
\label{smcv34}
\end{equation}
The equation for the total vorticity is therefore
\begin{equation}
\frac{\partial\overline{\omega}}{\partial t}+{\bf u}\cdot \nabla\overline{\omega}=\nabla\cdot \left\lbrace D({\bf r},t) \left\lbrack \nabla\overline{\omega}+(\Gamma_2^{f.g.}-\Gamma^2)\beta(t)\nabla\psi\right\rbrack\right\rbrace.
\label{smcv35}
\end{equation}
This equation coincides with Eq. (\ref{meppd2}) obtained for point vortices. We can also check that the equilibrium states of Eqs. (\ref{smcv33}) and (\ref{smcv35}) return Eqs. (\ref{smcv8b}) and (\ref{smcv12}) at leading order.

\section{Difference between the Joyce-Montgomery theory and the MRS theory}
\label{sec_diff}

It may be useful to re-emphasize the physical difference between the statistical theory of point vortices \cite{jm} and the statistical theory of continuous vorticity fields \cite{miller,rs}.

Let us first consider a continuous vorticity field $\omega({\bf r},t)$ whose evolution is described by the 2D Euler equation. If the initial state is unsteady or dynamically unstable, the fine-grained vorticity $\omega({\bf r},t)$ will undergo a complicated mixing process during which the system generates filaments at smaller and smaller scales. In this sense, there is no relaxation (the Euler equation is reversible). However, if we locally average over the filaments, the coarse-grained vorticity $\overline{\omega}({\bf r},t)$ will reach a steady state $\overline{\omega}({\bf r})$. This is because the evolution continues at a scale smaller than the resolution scale. The coarse-grained field  $\overline{\omega}({\bf r})$ is expected to be a stable steady state of the 2D Euler equation. If mixing is efficient, so that the  ergodicity assumption is fulfilled, the  coarse-grained field  $\overline{\omega}({\bf r})$ can be predicted by the MRS statistical theory (this theory also gives the fluctuations around this averaged field). It corresponds to the statistical equilibrium state of the 2D Euler equation. Since this equilibrium state is established very rapidly, after a few dynamical times, this process is called violent relaxation \cite{lb}. In reality, there is always some viscosity. If viscous effects are weak, the system first undergoes a violent relaxation towards a quasi stationary state, then decays on a longer timescale because of viscous effects.

We now consider a system of $N$ point vortices with circulation $\gamma\sim 1/N$
(for simplicity we assume that the point vortices have the same circulation but
the following arguments can be extended to more general cases). The {\it
discrete} vorticity field $\omega_d({\bf r},t)=\sum_i \gamma \delta({\bf r}-{\bf
r}_i(t))$ is a sum of $\delta$-functions. We note that $\omega_d({\bf r},t)$
satisfies the 2D Euler equation (called the Klimontovich equation in plasma
physics) for any $N$. If $N\gg 1$, then, in a first regime (collisionless
regime), the correlations between point vortices can be neglected and the
{\it smooth} vorticity field $\omega({\bf r},t)=\langle\omega_d({\bf
r},t)\rangle=\langle\sum_i \gamma \delta({\bf r}-{\bf r}_i)\rangle$ also
satisfies the 2D Euler equation (called the Vlasov equation in plasma physics)
\cite{pre,kinonsager}. In the limit $N\rightarrow +\infty$, the smooth
vorticity field $\omega({\bf r},t)$ satisfies the 2D Euler equation for all
times \cite{marchioro}.  Since the point vortex gas is (first) described by the
2D Euler equation, it  may undergo a process of violent relaxation towards a
quasi stationary state described by the MRS theory (assuming ergodicity). {\it
This quasi stationary state corresponds to the statistical equilibrium state of
the 2D Euler equation for a continuous vorticity field.} The relaxation time
towards that state is of the order of a few dynamical times. Then, in a second
regime (collisional regime), and since $N$ is necessarily finite, correlations
(also called distant collisions between point vortices) come into play and the
smooth vorticity field $\omega({\bf r},t)=\langle\omega_d({\bf
r},t)\rangle=\langle\sum_i \gamma \delta({\bf r}-{\bf r}_i)\rangle$ is not a
solution of the  2D Euler equation anymore. It is the solution of a kinetic
equation which includes a sort of collision term \cite{pre,kinonsager}. This
kinetic equation is expected to relax towards the Boltzmann distribution
corresponding to the Joyce-Montgomery theory. {\it This distribution corresponds
to the statistical equilibrium state of the discrete point vortex gas.} The
relaxation time towards that state depends on the number of point vortices and
increases rapidly with $N$. The scaling of the relaxation time with $N$ ($Nt_D$,
$N^2t_D$, $e^N t_D$?) is not firmly established. It is not even clear whether
the statistical equilibrium state is reached in all cases. Indeed, the evolution
may be non-ergodic. When $N\rightarrow +\infty$, the
collisional regime and the establishment of the Boltzmann distribution are
pushed
to infinitely long times so the domain of validity of the 2D Euler equation
(collisionless regime) becomes huge. In that case, only the QSS described by the
MRS theory can be observed.

These considerations of kinetic theory \cite{pre,kinonsager} make clear that the
MRS theory and the Joyce-Montgomery theory are fundamentally different (despite
their mathematical similarities) since they describe two very different regimes
with very different timescales: violent collisionless relaxation vs slow
collisional relaxation. It is not always clear in numerical simulations
of point vortices whether the observed ``equilibrium state'' corresponds to the
QSS distribution predicted by the MRS theory or to the Boltzmann distribution
predicted by Joyce and Montgomery. This important remark will be given more
consideration in future works. Note, finally, that for ``small'' $N$, the point
vortex gas may reach the Boltzmann distribution without passing through the
collisionless regime of violent relaxation.

{\it Remark:} This two-stages process for point vortices (violent collisionless
relaxation 
followed by slow collisional relaxation) is similar to what is known for stellar
systems \cite{houches,bt,paddy}. There is, however, a fundamental difference
between
stellar
systems and 2D vortices. Galaxies are basically made of a finite
number of stars and the Vlasov equation is an approximation of the
discrete dynamics of
stars for large $N$. Inversely, 2D flows are basically described by the 2D Euler
equation for a continuous vorticity field and the point vortex model with large
$N$ is an approximation of the continuous dynamics. Therefore, the physical
problematic is,
in some
sense, reversed.

\end{document}